\documentclass[journal,twocolumn]{IEEEtran}
\usepackage{epsfig,makeidx,color}
\usepackage{amsmath,amssymb,amsthm,bbm}
\usepackage{cite,graphicx}
\usepackage{enumerate}
\usepackage{hyperref}
\hypersetup{
        colorlinks = true,
        citecolor=blue,
}

\def\cL{{\cal L}}

\def\cX{{\cal X}}
\def\cY{{\cal Y}}

\def\rT{{\rm T}}

\def\uR{{\mathbb R}}

\def\uE{{\mathbb E}}

\DeclareMathOperator*{\argmin}{\arg\!\min}
\DeclareMathOperator*{\argmax}{\arg\!\max}

\newtheorem{mylemma}{\bf Lemma} 

\def\be{ \begin{equation} }
\def\ee{ \end{equation} }
\def\bea{ \begin{eqnarray} }
\def\eea{ \end{eqnarray} }
\def\bx{{\bf x}}
\def\by{{\bf y}}

\def\ba{{\bf a}}

\def\bk{{\bf k}}

\def\bp{{\bf p}}

\def\bw{{\bf w}}

\def\bA{{\bf A}}

\def\bI{{\bf I}}

\def\bK{{\bf K}}

\def\bR{{\bf R}}

\def\b0{{\bf 0}}

\def\bSigma{{\bf \Sigma}}

\def\cN{{\cal N}}

\def\cU{{\cal U}}

\def\sMSE{{\sf MSE}}
\def\sSSE{{\sf SSE}}

\ifCLASSOPTIONonecolumn
  \newcommand{\figwidth}{0.60\columnwidth}
\else
  \newcommand{\figwidth}{0.80\columnwidth}
\fi

\begin{document}

\title{Data-aided Sensing for Gaussian Process Regression in
IoT Systems}

\author{Jinho Choi
\thanks{The author is with
the School of Information Technology,
Deakin University, Geelong, VIC 3220, Australia
(e-mail: jinho.choi@deakin.edu.au).
This research was supported
by the Australian Government through the Australian Research
Council's Discovery Projects funding scheme (DP200100391).}}

\maketitle

\begin{abstract}
In this paper, 
for efficient data collection with limited bandwidth,
data-aided sensing is applied to 
Gaussian process regression that is used to learn data sets
collected from sensors
in Internet-of-Things systems.
We focus on the interpolation of sensors' measurements from 
a small number of measurements uploaded by a fraction of sensors
using 
Gaussian process regression with data-aided sensing. 
Thanks to active sensor selection,
it is shown that
Gaussian process regression with data-aided sensing
can provide a good estimate of a complete
data set compared to that with random selection.
With multichannel ALOHA, data-aided sensing is generalized
for distributed selective uploading
when sensors can have feedback of predictions of their measurements
so that each sensor can decide whether or not it uploads 
by comparing its measurement with the predicted one.
Numerical results show that modified
multichannel ALOHA with predictions can help improve
the performance of 
Gaussian process regression with data-aided sensing
compared to 
conventional multichannel ALOHA with equal uploading probability.
\end{abstract}

{\IEEEkeywords
Gaussian Process Regression;
Data-aided Sensing;
Active Learning}

\ifCLASSOPTIONonecolumn
\baselineskip 26pt
\fi

\section{Introduction} \label{S:Intro}

The Internet of Things (IoT) is a network
of things where devices and sensors
that are connected for
a number of applications 
in various areas including smart cities and factories
\cite{Gubbi13} \cite{Kim16}.
In general,
layered approaches are considered to build IoT systems,
where the bottom layer is usually
responsible for collecting information or data from devices or sensors
and the top layer is the application layer
\cite{Fuqaha15} \cite{Yaq17}.
Applications in the application layer
are to process data sets collected from
devices and sensors to produce desired outcomes.

To support the connectivity for the IoT,
a number of different approaches have
been proposed \cite{Ding_20Access}.
Among those, machine-type communication (MTC)
is one of the promising approaches 
that can support a large number of devices
through cellular systems.
In \cite{Mang16},
a deployment study of 
narrowband IoT (NB-IoT) \cite{3GPP_NBIoT}, 
which is one of MTC standards,
is carried out to support IoT applications over a large area.
A key advantage of cellular IoT is 
the infrastructure that allows
an number of IoT applications to utilize data 
sets collected from MTC-enabled sensors and devices
over a large geographical area (e.g., a region or country).

Collecting data sets from 
sensors or devices deployed in an area of interest
requires sensing to acquire local measurements or data 
and uploading to an access point (AP) 
in a wireless sensor network  (WSN)
or a base station (BS) in cellular IoT.
In general, energy efficient techniques to save sensors' energy
are widely studied \cite{Soua11} \cite{Abella19}.
However, bandwidth efficient techniques are not relatively well studied
(except for some specific cases, e.g.,
WSN for monitoring high frequency events in \cite{Bhuiyan13}), 
which may be crucial in MTC when there are a large number of
sensors \cite{Bockelmann18}.

Due to the presence of a large number of sensors
or devices in a cell, it is often challenging 
to collect data sets with limited bandwidth.
For example, suppose that a set of measurements 
can be periodically collected from sensors or devices
and stored in a cloud without any specific
applications. Then, since it may take a long time to collect
all the measurements,
measurements from certain devices may not be 
utilized for a while by any applications and become outdated
when they are eventually utilized.
Thus, while sensing and uploading can be considered separately,
they can be combined for efficient
data collection, which leads to
data-aided sensing (DAS) \cite{Choi19} \cite{Choi20_IoT}.
DAS is an iterative data collection scheme
where a BS is to collect data sets from devices or sensors
through multiple rounds.
In DAS, the BS is able to choose a set of sensors at each round
based on the data sets that are already available at the BS
from previous rounds for efficient data collection,
while the sensor (or device) selection criteria
can depend on specific applications.
As a result, with limited bandwidth, the BS is able to 
efficiently provide necessary information 
with a small number of measurements compared to random selection.

In \cite{Choi19} \cite{Choi20_IoT},
DAS has been studied with specific parametric models for measurements in
WSNs or IoT systems.
On the other hand, in this paper, 
without any specific model for measurements,
DAS is applied to Gaussian process regression (GPR), which is 
one of machine learning algorithms 
\cite{Williams96} \cite{Rasm06} \cite{Bishop06},
when GPR is used to learn a data set of sensors.
Note that various machine learning algorithms can be used
for IoT networks \cite{Wang20}.
In particular, with a partial set of measurements,
GPR is used for regression
so that an estimate of the measurements 
of the sensors that do not upload their measurements yet
can be obtained. To make GPR efficient
with a small number of sensors uploading their measurements,
DAS is applied for active sensor selection.
The resulting approach
can be seen as active learning 
\cite{Cohn96} in WSNs 
or IoT systems,
where data collection for learning can be actively
carried out through active sensor selection in DAS.
In this paper,
DAS is also generalized 
with selection criteria when multiple
sensors can upload at each round
through multiple channels using random access
(i.e., multichannel ALOHA \cite{Shen03} \cite{Chang15}).
The resulting DAS becomes distributed DAS
where distributed selective 
uploading takes place by multiple sensors
of different probabilities of uploading
that depend on the difference between
actual measurement and prediction.
Note that when some sensors go to sleeping mode to save energy \cite{Jurdak10},
the bandwidth can be wasted,
because the BS cannot receive any measurements from them in 
conventional DAS.
On the other hand, in distributed DAS, since only
active sensors can upload, it can be more efficient 
than conventional DAS when a fraction of sensors are dormant.

The rest of the paper is organized
as follows. In Section~\ref{S:BG},
the system model is presented with an overview
of GPR that allows to perform regression 
with a subset of measurements
uploaded by a fraction of sensors.
DAS is applied to GPR in Section~\ref{S:DAS}
and application dependent criteria for sensor selection
is introduced in Section~\ref{S:ADC}.
In Section~\ref{S:Dist}, a distributed
updating approach for DAS is studied
with multichannel ALOHA. We present
numerical results in Section~\ref{S:Sim}
and finally conclude the paper in 
Section~\ref{S:Con} with some remarks.

\subsubsection*{Notation}
Matrices and vectors are denoted by upper- and lower-case
boldface letters, respectively.
For a vector $\bx$, $||\bx||$ stands for the 2-norm.
The superscript $\rT$ 
denotes the transpose and
the superscript $c$ represents the complement of a set.
$\uE[\cdot]$
and ${\rm Cov}(\cdot)$
denote the statistical expectation and covariance, respectively.
$\cN(\ba, \bR)$
represents the distribution of
Gaussian random vectors with mean vector $\ba$ and
covariance matrix $\bR$.
${\rm Unif}(a,b)$ stands for
the uniform distribution over $[a,b)$, where $b > a$.

\begin{table}[tbh]
        \label{TBL:1}
\caption{Glossary of key symbols.}
\begin{tabular}{ll}
$L$: & Number of sensors \cr
$\bx_l$: & Location of sensor $l$ \cr
$\cX$: & Set of the locations of all $L$ sensors \cr
$y_l$: & Measurement of sensor $l$ \cr
$\by(t)$: & Set of measurements up to round $t$ \cr
$\cY$: & Set of the measurements of all $L$ sensors \cr
$m(\bx_l)$: & Mean measurement at location $\bx_l$ \cr
$B$: & Number of multiple access channels \cr
$Q$: & Number of selected sensors per round 
with $B$ channels \cr
$p_{\rm slee}$: & Probability that a sensor in sleep mode \cr
\end{tabular}
\end{table}

\section{Background} \label{S:BG}

In this section, we present the system model
and an overview of GPR.

\subsection{Sensors and Measurements}

\begin{figure}[thb]
\begin{center}
\includegraphics[width=\figwidth]{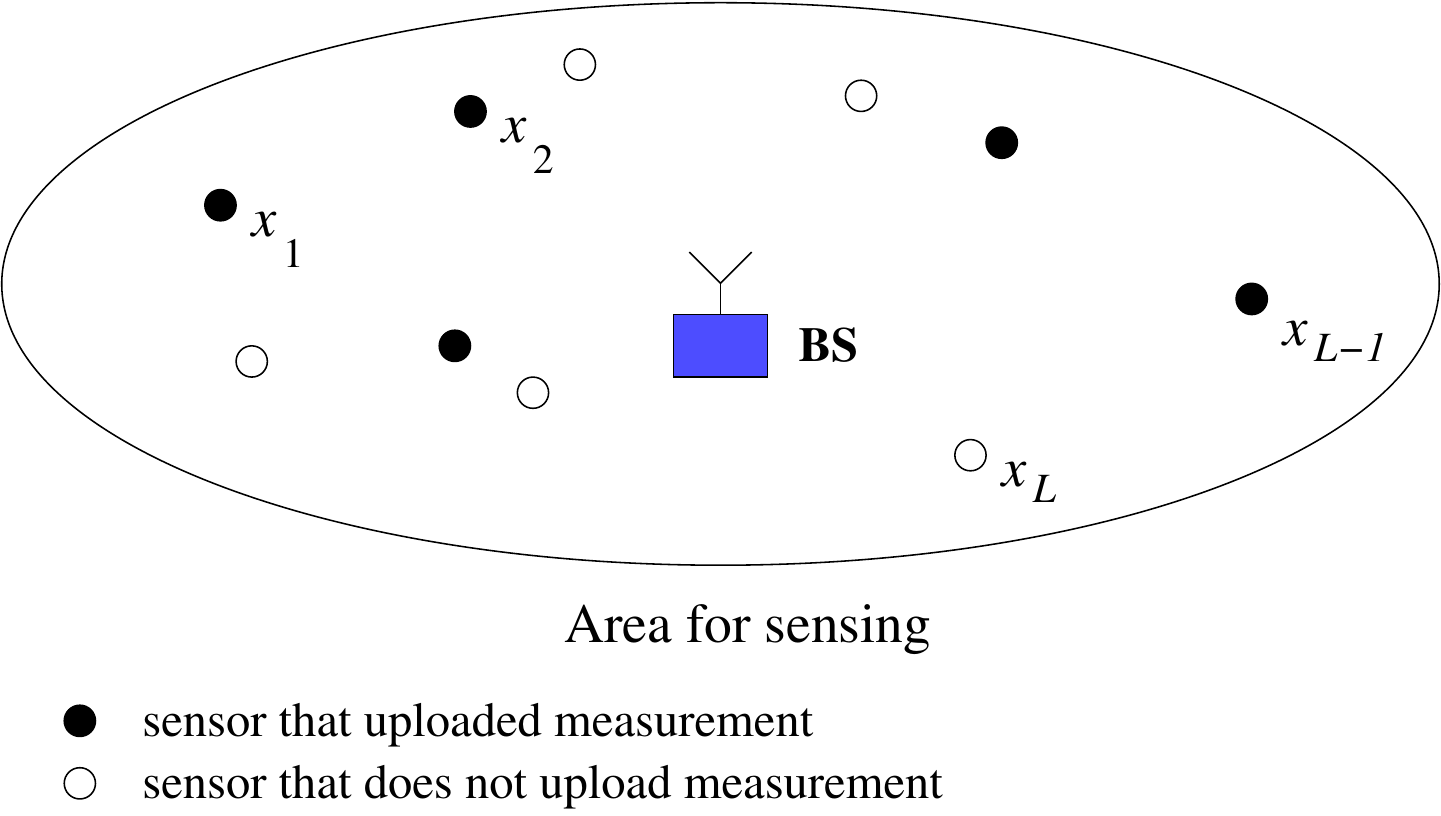}
\end{center}
\caption{An illustration of an IoT system
consisting of $L$ sensors and a BS.}
        \label{Fig:wsn}
\end{figure}

Suppose that there are $L$ sensors and each sensor's
location, denoted by $\bx_l$, is known 
by a BS that is also deployed with sensors
over a certain area for a cellular IoT system (or WSN)
as illustrated in Fig.~\ref{Fig:wsn}.
Here, the subscript, $l$,
is used to denote the index of sensors.
The measurement of sensor $l$ is denoted by $y_l$. 
Thus, the complete data set
is the set of all measurements, which is denoted by 
$\cY = \{y_1, \ldots, y_L\}$.
For simplicity, we assume that $y_l$ is a scalar,
although it can be a vector. For example, $y_l$
can represent the temperature at location $\bx_l$
when an IoT system is used for environmental monitoring.
In addition, since $\bx_l$ 
represents the location
of sensor $l$, it can be a scalar in a one-dimensional space
or a 2-dimensional vector in 
a two-dimensional Cartesian coordinate system
(which might be typical in most WSNs).

If the BS can receive the measurements from all the sensors,
it has the complete data
set, $\cY$, that shall be available for a number of 
applications in IoT systems \cite{Fuqaha15}.
However, when $L$ is sufficiently large,
with a limited bandwidth, it may take a long time to make 
$\cY$ available. As a result, when the uploading time is limited,
it is desirable
to find an estimate of $\cY$ using measurements
from a small number of sensors.
That is, as shown in Fig.~\ref{Fig:wsn},
with a partial data set uploaded by a fraction of sensors 
(represented by black dots), the BS needs to 
perform interpolation to estimate the complete data set, $\cY$.
Fortunately, provided that the measurements of
closely located sensors are highly correlated,
it is possible to have a good estimate
of $\cY$ from measurements of a fraction of sensors
using ML algorithms. Throughout the paper, we assume that the BS
has sufficient computing power to perform ML algorithms.
In this sense, the BS can also be seen as a server that
performs ML algorithms to learn data sets.


\subsection{GPR}

In this subsection, we present an overview of 
one of ML approaches, GPR \cite{Williams96}, to learn data.
Since GPR is a nonparametric regression approach, 
we will use it to estimate $\cY$ using a subset of $\cY$
without any specific parametric model. 

Let $\bx_l \in \cX$ be the $l$th input location,
where $\cX$ represents the set of all possible locations
(in the time or space\footnote{In this paper,
we consider the space domain as GPR is used
to interpolate measurements of sensors that
are deployed over a certain area.} 
domain).
In addition, denote by $y(\bx_l)$ 
the measurement or data at $\bx_l$, i.e., $y_l = y(\bx_l)$.
It is assumed that 
\be
y(\bx_l) \sim \cN(m (\bx_l), \sigma^2),
\ee
where 
$m(\bx_l)$ denotes the mean of the measurement at location $\bx_l$ and
$\sigma^2$ represents the variance of noisy
observation. Furthermore, $m(\bx)$ is assumed
to be Gaussian that has the following 
distribution:
\be
m(\bx) \sim \cN (0, k_\theta),
\ee
where $k_\theta$ represents the kernel that is parameterized by
$\theta$ 
 \cite{Williams96}\cite{Rasm06} \cite{Bishop06}.

Suppose that
the measurements at certain locations, which are
denoted by $\bx_1, \ldots, \bx_N$, are available. That is,
$y(\bx)$ is available for $\bx \in \cX_0  = \{\bx_1, \ldots, \bx_N\}
\subset \cX$. 
For convenience,
let $\by_0 = [y(\bx_1) \ \ldots \ y(\bx_N)]^\rT$.
In addition,
let $\cX_1 = \{\bx_{N+1}, \ldots, \bx_{N+M}\} \subset \cX$
be the test set, which is a set of certain locations of interest. 
Using GPR, although
the measurements associated with $\cX_1$ are not available,
it is possible to find their estimates
via the distribution of their mean values,
which is the following conditional distribution:
\be
f(m (\cX_1)\,|\, \by_0, \cX_0, \cX_1),
\ee
where $m(\cX_1) = [m(\bx_{N+1}) \ \ldots \ m(\bx_{N+M})]^\rT$.
Under the Gaussian process assumption,
the joint distribution of $\by_0$ and $m(\cX_1)$
can be given by
\be
\left[
\begin{array}{c}
\by_0 \cr
m(\cX_1) \cr
\end{array}
\right]
\sim \cN\left(\b0,
\left[
\begin{array}{cc}
\bK_\theta (\cX_0, \cX_0) + \sigma^2 & \bK_\theta (\cX_0, \cX_1) \cr
\bK_\theta (\cX_1, \cX_0) & \bK_\theta (\cX_1, \cX_1) \cr
\end{array}
\right]
\right),
	\label{EQ:JD}
\ee
where $\bK_\theta (\cX_i, \cX_j)$
represents 
the covariance matrix comprising
the covariance between an element of $\cX_i$ and
another element of $\cX_j$, where $i, j \in \{0,1\}$.
For example, 
$[\bK_\theta (\cX_0, \cX_1)]_{l, m} = 
k_\theta (\bx_l, \bx_{l+m})$,
$l = 1, \ldots, N$ and $m = 1, \ldots, M$.

From \eqref{EQ:JD},
the conditional distribution of $m(\cX_1)$, which
is also Gaussian, has the mean and covariance matrix as follows:
\begin{align}
\uE[m(\cX_1)\,|\, \by_0]
& = 
\bK_\theta (\cX_1, \cX_0) 
\left(\bK_\theta (\cX_0, \cX_0)  + \sigma^2 \bI\right)^{-1} \by_0 \cr
{\rm Cov}(m(\cX_1)\,|\, \by_0)
& = \bK_\theta (\cX_1, \cX_1) 
- \bK_\theta (\cX_1, \cX_0)  \cr
& \times
\left(\bK_\theta (\cX_0, \cX_0)  + \sigma^2 \bI\right)^{-1} 
\bK_\theta (\cX_0, \cX_1) .
	\label{EQ:MMC}
\end{align}
In \eqref{EQ:MMC}, 
$\uE[m(\cX_1)\,|\, \by_0]$ can be
seen as an estimate 
or prediction\footnote{Since we consider
the space domain in a WSN,
$\uE[m(\cX_1)\,|\, \by_0]$ is an estimate
rather than an prediction as the measurements
of the sensors are time-invariant. However, we use 
the terms, estimation,
interpolation, and prediction, interchangeably in this paper.}
of the measurements of the
sensors associated with $\cX_1$ for given
available measurements $\by_0$,
and 
the conditional covariance matrix,
${\rm Cov}(m(\cX_1)\,|\, \by_0)$, can be used to
see the estimation error or uncertainty
of $\uE[m(\cX_1)\,|\, \by_0]$.

\section{Data-aided Sensing for GPR}	\label{S:DAS}

Due to limited bandwidth for uplink transmissions,
the BS may need to collect measurements from sensors 
through a number of rounds. Then, there can be a 
pre-determined order for sensors' uploading,
e.g., the increasing order in terms of the index, $l$.
While any order can be considered,
there could be a certain order
that allows the BS to produce a good prediction of the values of
the sensors that do not upload their measurements
yet using the measurements from the sensors that already uploaded.
The resulting approach is referred to as
DAS in \cite{Choi19} \cite{Choi20_IoT}.
In this section, DAS is applied to GPR
where regression can be performed with
\emph{active sensor selection} so that
a good estimate of the sensors' complete data set, i.e., $\cY$,
is available with
a relatively small number of sensors.

For simplicity, suppose that one sensor can upload
its measurement at each round.
Denote by $l(t)$ the index
of the sensor that uploads its measurement at round $t$, $t \ge 1$.
In addition, let
$\cL(t)$ be the index set of the sensors that
have uploaded their measurements up to round $t$, 
i.e., $\cL (t) = \{l(1), \ldots, l(t)\}$.
For convenience, let $\cL(0) = \emptyset$.
In addition, let
\begin{align*}
\by (t)\!: & \ \mbox{the measurements corresponding to $\cL(t)$} \cr
\cX_0 (t)\!: & \ \mbox{the set of locations of sensors corresponding to $\cL(t)$} \cr
\cX_1 (t)\!: & \ \mbox{the set of locations of sensors 
corresponding to $\cL(t)^c$}.
\end{align*}

Since $\by(t) \in \uR^t$ becomes the vector of the
measurements that are available at the BS up to round $t$,
the BS can use GPR to
find the conditional mean of $m(\cX_1(t))$ 
(for given $\by(t)$)
as an estimate
of the measurements of the sensors that do not upload
yet (i.e., the sensors associated with $\cL(t)^c$).
That is, at round $t$, an estimate of 
the complete data set, $\cY$,
can be obtained from the partial data set associated
with $\cX_0(t)$, i.e., $\by(t)$. The prediction 
for the measurements associated with $\cX_1(t)$,
$\uE[m(\cX_1(t))\,|\, \by(t)]$,
using GPR is given by 
\be
\hat \cY (t) = 
\left[ \begin{array}{c}
\by (t) \cr
\uE[m(\cX_1(t))\,|\, \by(t)] \cr
\end{array} \right],
\ee
where
$\hat \cY(t)$ denotes the estimate of $\cY$ at round $t$.
The 
estimation error is dependent on the 
conditional covariance of $m(\cX_1(t))$,
i.e., ${\rm Cov} (m(\cX_1 (t) \,|\, \by (t))$.
In particular, the 
mean squared error
(MSE) of the estimate is given by
\begin{align}
& \uE[||m(\cX_1 (t)) - \uE[ m(\cX_1 (t))\,|\, \by(t)]||^2] \cr
& \quad = {\rm Tr} \left({\rm Cov} (m(\cX_1 (t) \,|\, \by(t))\right).
\end{align}

At the next round, i.e., round $t+1$,
a sensor can be selected by the BS to upload its measurement and
the selection can be based on the minimization of MSE.
To this end, with a certain $l \in \cL (t)^c$,
consider the joint distribution of $\by(t)$ and $m(\bx_l)$ that
has the following covariance matrix:
\begin{align}
{\rm Cov} 
\left(\left[
\begin{array}{c}
\by (t) \cr m(\bx_l) \cr
\end{array}
\right] \right)
= 
\left[
\begin{array}{cc}
\bK_\theta (\cX_0, \cX_0) + \sigma^2 & \bk_\theta(\cX_0, \bx_l) \cr
\bk_\theta(\cX_0, \bx_l)^\rT & k_\theta(\bx_l, \bx_l) \cr
\end{array}
\right].
\end{align}
Then, the conditional mean of $m(\bx_l)$ for given $\by(t)$
becomes
\be
m(\bx_l\,|\, \by (t))
= 
\bk_\theta (\cX_{0}, \bx_{l})^\rT
\left(\bK_\theta (\cX_0, \cX_0)  + \sigma^2 \bI\right)^{-1} \by(t)
	\label{EQ:gpr1}
\ee
and the conditional variance is given by
\begin{align}
& {\rm Cov} ( m(\bx_l \,|\, \by(t)) ) = k_\theta (\bx_l, \bx_l) \cr
& - 
\bk_\theta (\cX_{0}, \bx_{l})^\rT
\left(\bK_\theta (\cX_0, \cX_0)  + \sigma^2 \bI\right)^{-1} 
\bk_\theta (\cX_{0}, \bx_{l}).
	\label{EQ:gpr2}
\end{align}

\begin{mylemma}
Among all $l \in \cL(t)^c$, the next sensor that can 
minimize the MSE of the estimate of $m (\cX_1 (t))$
is the sensor associated with the maximum variance.
As a result, the selection criterion to choose the next
sensor in DAS is as follows:
\be
l^* (t+1) 
= \argmax_{l \in \cL^c (t)} 
{\rm Cov} ( m(\bx_{l} \,|\, \by(t) )).
	\label{EQ:lt1}
\ee
\end{mylemma}
\begin{IEEEproof}
The sum of the conditional variances of $\bx_l$ for all $l \in \cL (t)^c$
is also the MSE of the estimate of $m (\cX_1 (t))$,
$m(\cX_1 (t) |\by (t))$,
as follows:
\begin{align}
\sMSE (t) 
& =  \uE[||m(\cX_1(t)) - \uE[m(\cX_1 (t) \,|\, \by(t) ||^2] \cr
& = \sum_{l \in \cL (t)^c} {\rm Cov} ( m(\bx_l \,|\, \by(t)) ).
	\label{EQ:MSEt}
\end{align}
Suppose that $l \in \cL(t)^c$ be the index of the next sensor 
so that $\cL (t+1) = \{\cL(t), l\}$
and $\cL (t+1)^c = \cL(t) \setminus \{l\}$. Then,
we have
$$
\sMSE (t) = 
{\rm Cov} ( m(\bx_l \,|\, \by(t)) )
+\sum_{l^\prime \in \cL (t)^c \setminus \{l\}} 
{\rm Cov} ( m(\bx_{l^\prime} \,|\, \by(t)) ),
$$
where the second term on the right-hand side 
becomes the MSE at round $t+1$.
Since $\sMSE(t)$ is independent of the selection of $l$,
the minimization of the MSE at round $t+1$,
i.e.,
$\min_l \sum_{l^\prime \in \cL (t)^c \setminus \{l\}} 
{\rm Cov} ( m(\bx_{l^\prime} \,|\, \by(t)) )$,
is equivalent to the maximization of 
${\rm Cov} ( m(\bx_l \,|\, \by(t)) )$,
i.e., $\max_l {\rm Cov} ( m(\bx_l \,|\, \by(t)) )$,
which leads to the sensor selection 
criterion in \eqref{EQ:lt1} for DAS.
\end{IEEEproof}

Based on \eqref{EQ:lt1},
DAS can be applied to GPR for
a good estimate of the complete data set, $\cY$,
with a small number of the sensors that upload their measurements.
There are a few remarks as follows.
\begin{itemize}

\item We have assumed that $\cX_1 (t)$
is the set of the locations of the sensors that
do not upload their measurements yet.
It is also possible to define $\cX_1 (t)$ differently so that
the measurements at locations without sensors can be estimated/predicted.
In particular, suppose that it is required to interpolate 
\emph{virtual} measurements at some locations' where
no sensors exist. Let $\tilde \cX_1 = \{\tilde \bx_1,  
\ldots, \tilde \bx_M\}$, where $\tilde \bx_m$
represents the $m$th location of interest with
$\tilde \bx_m \notin \cX$.
Then, DAS can be used for interpolation,
and the next sensor to upload
its measurement 
to minimize the MSE 
can be decided as follows:
\be
l^* (t+1) = \argmax_{l \in \cL^c (t)} 
{\rm Cov} (m(\tilde \cX_1)\,|\, \by (t), y(\bx_l)).
	\label{EQ:lt1}
\ee
Note that the complexity of DAS is proportional to that of GPR 
as shown in \eqref{EQ:lt1}, which is mainly dependent
on the complexity of matrix inversion
in \eqref{EQ:gpr1} and \eqref{EQ:gpr2} \cite{Williams96}.

\item
Like Gaussian DAS in \cite{Choi20_IoT},
\eqref{EQ:lt1} does not depend on the actual measurements,
$\cY$, or available measurements, $\by (t)$.  
In other words,
the order of sensors to upload 
does not depend on the existing data set, $\by (t)$.
However, as will be shown in Section~\ref{S:ADC},
there are also cases that the next sensor to upload
its measurement depends on the existing data set
or the current prediction of $\cY$.

\item 
Although we only consider the case
that all the measurements remain unchanged
over a number of uploading rounds,
they can be time-varying. Thus, to estimate $\cY$
within a limited time (for some real-time applications),
DAS is more desirable to efficiently upload 
sensors' measurements.
DAS for time-varying models subject to delay constraints
might be a further research topic to be investigated
in the future.
\end{itemize}

\section{Application-Dependent Criteria}	\label{S:ADC}

In this section, we generalize DAS 
for active sensor selection to the case that
multiple applications need an estimate of $\cY$
for different purposes.

Suppose that there are a number of different 
applications that use the measurements collected by the BS.
Thus, the BS is to provide 
an estimate of $\cY$ using the measurements
up to round $t$, $\by (t)$, i.e., $\hat \cY(t)$.
Then, the conditional distribution of $\hat  \cY(t)$
becomes
\be
\hat \cY(t) \,|\, \by(t) \sim 
\cN
\left(
\left[ \begin{array}{c}
\by (t) \cr
m(\cX_1 (t)\,|\, \by (t)) \cr
\end{array} \right],
\bSigma_{\by} (t) 
\right),
\ee
where
\be
\bSigma_{\by} (t) = 
\left[
\begin{array}{cc}
\b0 & \b0 \cr
\b0 & {\rm Cov} (m(\cX_1(t))\,|\, \by (t)) \cr
\end{array}
\right].
\ee

For example, suppose that $y_l = y(\bx_l)$ 
is the temperature at a location of 
$\bx_l$.
Then, an application\footnote{In this paper, applications
mean \emph{functions} of $\cY$ or $\cY(t)$ the
produce outcomes for applications in the application layer
of an IoT system.}
that is to provide the real-time mean 
temperature over the area where sensors are deployed 
can have the
following estimate of the mean temperature:
$$
\hat \mu (t) = \bw^\rT \hat \cY(t),
$$
where $\bw = [\frac{1}{L} \ \ldots \ \frac{1}{L}]^\rT$.
In general, the output of an application can be given by
\be
Z_q (t) = \bw_q^\rT \hat \cY(t),
\ee
where $\bw_q$ represents the weighting vector
for the $q$th \emph{linear} application. Here,
by a linear application, we mean
an application that provides an output as a linear function
of $\hat \cY(t)$.
For a linear application,
the MSE of the output can be given by
\be
\sMSE_q (t) = \bw_q^\rT \bSigma_{\by} (t) \bw_q.
\ee

Suppose that the $l$th sensor, $l \in \cL(t)^c$,
is chosen at round $t+1$. The conditional covariance 
matrix can be updated as 
follows:
\be
\bSigma_{\by, l} (t+1) =
\left[
\begin{array}{cc}
\b0 & \b0 \cr
\b0 & {\rm Cov} (m(\cX_1(t))\,|\, \by (t), y(\bx_l)) \cr
\end{array}
\right].
\ee
Then, for the $q$th application
to minimize the MSE, the sensor to upload its
measurement at round $t+1$ can be chosen as follows:
\be
l_q (t+1) = \argmin_{l \in \cL (t)^c} \sMSE_{q,l} (t+1),
\ee
where
\be
\sMSE_{q, l} (t+1) = \bw_q^\rT \bSigma_{\by, l} (t) \bw_q.
\ee
If there are $Q$ applications, there can be up to $Q$ different sensors
that are to be selected at round $t+1$,
i.e., $\{l_1 (t+1), \ldots, l_Q (t+1)\}$.
Alternatively, 
one sensor can be chosen at a time 
(or at each round) according to the weighted sum of MSEs
as follows:
\be
l (t+1) = \argmin_{l \in \cL (t)^c} 
\sum_q \beta_q \sMSE_{q,l} (t+1),
\ee
where $\beta_q > 0$ is the weight for the $q$th application.

There can be nonlinear applications. For example,
if an application is to provide the maximum
temperature over the area, the output is
\be
Z = \max_l [\hat \cY (t)]_l.
\ee
Let $l^*(t) = \argmax_l [\hat \cY (t)]_l$.
If $l^*(t) \in \cL(t)$,
the maximum temperature is one of reported measurements up to round $t$.
Otherwise, it becomes one of the
predicted measurements of the sensors that do not upload yet. In this case,
the application may wish to know its actual measurement.
Thus, the index of the sensor at round $t+1$
can be given  by $l(t+1) = l^*(t) \in \cL (t)^c$.

Consequently, 
due to the presence of 
multiple applications (either linear or nonlinear
applications), we can assume that 
there are multiple candidates sensors 
(say $Q$ sensors) to upload their measurements 
at each round.
If the bandwidth of uplink channels is sufficiently wide,
at each round, $Q$ different channels can be
used to upload the measurements from $Q$ sensors simultaneously.
However, if the bandwidth is limited, 
a fraction of $Q$ sensors may be able to send their measurements.
In addition, there can be some sensors with their 
actual measurements
that are sufficiently close to their predictions, 
in which case their uploading may result in waste of resource
(i.e., uplink channels).
In this case, the sensors may not need to upload their measurements.
Thus, in DAS with multiple applications,
it would be necessary to consider efficient uploading approaches
due to limited bandwidth.

\section{Distributed Updating
using Multichannel Random Access with Predictions}
\label{S:Dist}

In this section, we study a distributed
approach to upload sensors' measurements using multichannel random access.

\subsection{Motivation}

As discussed in Section~\ref{S:ADC},
there can be $Q \ (\ge 1)$ sensors that are selected to upload their
measurements at each round. Thus, for simultaneous
uplink transmissions,
multiple access channels can be used, say $B$ channels,
where a sensor can upload its measurement through a dedicated channel
at each round based on coordination by the BS
(i.e., the BS can assign $B$ 
uplink channels to $B$ (out of $Q$) sensors).
As a result, up to $B$ sensors
can upload their measurements at each round.
However, there can be some drawbacks as follows:
\emph{i)} some sensors may not have measurements
due to various reasons (e.g., a sensor may be in dormant mode
and so that it is unable to respond) \cite{Choi20_IoT}; 
\emph{ii)} some sensors have measurements that are sufficiently
close to their predictions at the BS and their uploading
leads to waste of channel resource;
\emph{iii)} if $Q > B$, a single round is not sufficient.
Thus, to address
the those drawbacks, we consider 
a random access scheme, i.e.,
multichannel ALOHA with feedback
of predictions,  which will be discussed in the next subsection,
rather than uploading coordinated by the BS.

Note that since multichannel ALOHA is employed for
MTC \cite{Galinina13}\cite{Chang15} \cite{Choi16CL},
DAS with multichannel ALOHA could be easily adopted
with MTC in cellular IoT.

\subsection{Distributed Updating using Multichannel ALOHA}

Suppose that there are $B$ multiple access channels
for simultaneous uplink transmissions from multiple sensors.
We assume that the BS can choose $Q$ sensors for uploading
before round $t+1$, where $Q \ge B$. 
The index set of $Q$ sensors
is denoted by 
$$
\cL_Q (t+1) = \{l_1 (t+1), \ldots, l_Q(t+1)\},
$$
where $l_q (t+1)$ represents the index of the $q$th
candidate sensor at round $t+1$.
Suppose that each of $Q$ sensors can decide
whether or not it uploads its measurement
independently with a certainly probability of uploading,
which is denoted by $p_{\rm up}$.
If a sensor decides to upload, 
which is referred to as an active sensor,
it can choose
one of $B$ channels uniformly at random.
For convenience, let $K$ be the number of the sensors
that decide to upload, which is a binomial random variable.
Among $K$ active sensors, there can be multiple sensors
that choose the same channel among $B$ channels, which
results in collision. If collision happens, it is assumed
that the BS cannot receive any measurements 
from the associated sensors.
Since the probability
that an active sensor can successfully transmit its measurement
without collision is
$\left(1 - \frac{1}{B} \right)^{K-1}$,
the average number of the sensors
that can successfully upload their measurements
without collisions is given by
\begin{align}
S & = \uE\left[K \left(1 - \frac{1}{B} \right)^{K-1} \right] \cr
& = \sum_{k=0}^Q k\left(1 - \frac{1}{B} \right)^{k-1}
\binom{Q}{k} p_{\rm up}^k (1- p_{\rm up})^{Q-k}. 
\end{align}
After some manipulations, it can be shown that
\be
S = p_{\rm up} Q \left(1 - 
\frac{p_{\rm up}}{B} \right)^{Q-1}.
\ee
It is straightforward to show that $S$ is maximized
when 
\be
p_{\rm up} = \frac{B}{Q}
\ee
and
the maximum $S$ is
\be
S_{\rm max} = B \left(1  - \frac{1}{Q} \right)^{Q-1}
\approx B e^{-1}.
	\label{EQ:Smax}
\ee

Note that as mentioned earlier, 
suppose that some of sensors may
not be able to upload their measurements
with a probability of $p_{\rm sleep}$,
In this case, for given $B$ channels,
$Q$ can be decided as
\be
Q = \frac{B}{1-p_{\rm sleep}},
\ee
so that the maximum $S$ can be achieved
where $1- p_{\rm sleep}$ becomes
effectively equal to the probability of uploading,
$p_{\rm up}$.

Furthermore, as mentioned earlier,
since the measurements
of some of $Q$ sensors are close to predicted values
by the BS, they may not need to upload.
To modify multichannel ALOHA with the feedback of predictions,
suppose that the BS can send the signals through downlink channel
to the $Q$ candidate sensors with the predicted
values of their measurements that
are given by
\be 
m_q (t) = m (\bx_q \,|\, \by(t)), \ q \in \cL_Q (t+1).
\ee
Then, at sensor $q \in \cL_Q (t+1)$, 
with $m_q (t)$,
the prediction error can be found as
\be
e_q = m_q (t) - y(\bx_q).
	\label{EQ:e_q}
\ee
For convenience, the sensor associated with
$l_q (t+1)$ is referred to as the $q$th sensor in this section
and omit the round index $t+1$.
In addition, let $p_q$ be the probability of uploading
of the $q$th sensor, which is now different
for each sensor in the modified multichannel ALOHA.
For given $p_q$, the $q$th sensor
can generate a binary random variable, denoted by $V_q$,
as follows:
\be
V_q = \left\{
\begin{array}{ll}
1, & \mbox{w.p. $p_q$} \cr
0, & \mbox{w.p. $1- p_q$} \cr
\end{array}
\right.
\ee
If $V_q = 1$, the $q$th sensor chooses one of $B$
channels uniformly at random and transmits its measurement
as in conventional multichannel ALOHA \cite{Shen03} \cite{Chang15}.
Otherwise, the $q$th sensor does not transmit.

Recall that $K$ is the number of active sensors,
which can be written as
\be
K = \sum_{q=1}^Q V_q \le Q.
\ee
The probability that the $q$th sensor 
(regardless of its activity)
eventually uploads its measurement can be given by
\begin{align}
s_q & = 0 \times \Pr(V_q = 0) \cr
& \quad + 
\uE
\left[\left(1 - \frac{1}{B} \right)^{K-1}\,|\, V_q = 1 \right]
\times \Pr(V_q = 1) \cr
& = \uE[V_q \psi_B^{\sum_{t \ne q} V_t}],
\end{align}
where $\psi_B =1 - \frac{1}{B}$.
Since the $V_q$'s are independent, we have
\begin{align}
s_q & = \uE[V_q] \uE[ \psi_B^{\sum_{t \ne q} V_t}]
= p_q \uE[ \prod_{t \ne q} \psi_B^{V_t}] \cr
& = p_q  \prod_{t \ne q} \uE[\psi_B^{V_t}] 
= p_q  \prod_{t \ne q} (1 - p_t + \psi_B p_t) \cr
& = {p_q}
\prod_{t \ne q} \left(1 - \frac{p_t}{B} \right).
	\label{EQ:s_q}
\end{align}
From \eqref{EQ:s_q}, we have $s_q \le p_q$.
In addition, if $B$ is sufficiently large,
we can have
\be
s_q = p_q \prod_{t \ne q} \left(1 - \frac{p_t}{B} \right)
\approx p_q e^{ - \sum_{t \ne q} \frac{p_t}{B} } .
\ee
We can further show that
\begin{align*}
\sum_q s_q 
& \approx \sum_q p_q e^{ - \sum_{t \ne q} \frac{p_t}{B} } \cr
& \approx \sum_q p_q e^{ - \frac{\sum_t p_t}{B} } 
 = P e^{- \frac{P}{B} } = B \frac{P}{B} e^{- \frac{P}{B} } \le B e^{-1} ,
\end{align*}
where $P = \sum_q p_q$.
In general, $\sum_q s_q$ is maximized when $\sum_q p_q = B$.
Furthermore, to minimize collisions, we can have
the following constraint on $p_q$:
\be
\sum_{q=1}^Q p_q \le B.
\ee

At the BS, if sensor $q$ transmits without
collision, the measurement error
becomes 0. Otherwise, the measurement error
is $m_q(t) - y(\bx_q)$.
Thus, the sum squared error (SSE) per
round at the BS is
\be
\sSSE = \sum_{q=1}^Q e_q^2 (1- s_q),
\ee
which can be minimized by 
\emph{distributed selective 
uploading} by $Q$ sensors that can decide $s_q$ or $p_q$
according to $e_q^2$.
In \eqref{EQ:s_q}, assuming that $\sum_{q} p_q \approx B$,
we have $s_q \approx p_q e^{-1}$, and since $1 - x< e^{-x}$ for 
a positive $x \ll 1$,
\be
\sSSE \approx \sum_{q=1}^Q e_q^2 (1- e^{-1} p_q)
\le \sum_{q=1}^Q e_q^2 \exp\left(- e^{-1} p_q \right).
	\label{EQ:MU}
\ee
Thus, to minimize the SSE, 
using the upper bound in
\eqref{EQ:MU}, the following optimization can be considered:
\begin{eqnarray}
& \min_{\{p_q\}} \sum_{q=1}^Q e_q^2 \exp\left(- e^{-1} p_q \right) & \cr
& \mbox{subject to} \ \sum_{q=1}^Q p_q \le B. &
	\label{EQ:O1}
\end{eqnarray}
For a given Lagrange multiplier $\lambda$,
each sensor can find $p_q$ that minimizes
its cost function in a distributed manner as follows:
\be
p_q (\lambda) =  [ e(\ln e_q^2 - \psi)]_0^1,
	\label{EQ:p_ql}
\ee
where
$[x]_a^b = \min\{b, \max\{x,a\}\}$ for $b \ge a$
and $\psi = \ln(\lambda) +1$ (or $\lambda = e^{\psi -1}$).
Clearly, \eqref{EQ:p_ql}
shows that the probability of uploading, $p_q$,
can be decided at sensor $q$ provided that $\lambda$ is given.

At the BS, using dual ascent \cite{Boyd11}
$\psi$ can be updated as follows:
\be
\psi(t+1) = \psi(t) + \mu
\left(K(t)- B\right),
	\label{EQ:ll}
\ee
where $\mu > 0$ is the step-size and $K (t)$
is the number of active devices at round $t$,
which is an estimate of $\sum_q p_q$.
Note that a similar distributed approach has been used
for federated learning with multichannel ALOHA 
in \cite{Choi20_WCL}.  Together with the feedback of predicted values,
$\{m_q (t)\}$,  the BS can also broadcast
$\psi(t)$ so that the sensors can decide
the probability of uploading according to \eqref{EQ:p_ql}.
Note that at each round, since the SSEs are different,
we cannot expect that $\psi(t)$ converges to a certain
constant. 

\section{Numerical Results}	\label{S:Sim}

In this section, we present numerical results
to see the performance of GPR with DAS 
for different models for $\{\bx_l, y_l\}$.
For simulations, the following kernel function is used
$$
k_\theta (\bx_l, \bx_{l^\prime})
= \exp\left( - \frac{1}{2} ||\bx_l -\bx_{l^\prime} ||^2 \right).
$$

\subsection{Results for GPR with DAS}

In this subsection,
we present numerical results for GPR
with random selection and DAS. 
Random selection is to choose one of the sensors
associated with $\cX_1 (t)$ uniformly at random at each round.
It is assumed that at each round,
only one sensor uploads its measurement.
To see how DAS works, only one run is considered
with $L$ randomly located sensors 
and the performance measure is the MSE in \eqref{EQ:MSEt}.
In addition, a sufficiently large $L$ is assumed so that
the MSE after $L$ rounds can be sufficiently small
(i.e., approaching 0).

\begin{figure}[thb]
\begin{center}
\includegraphics[width=\figwidth]{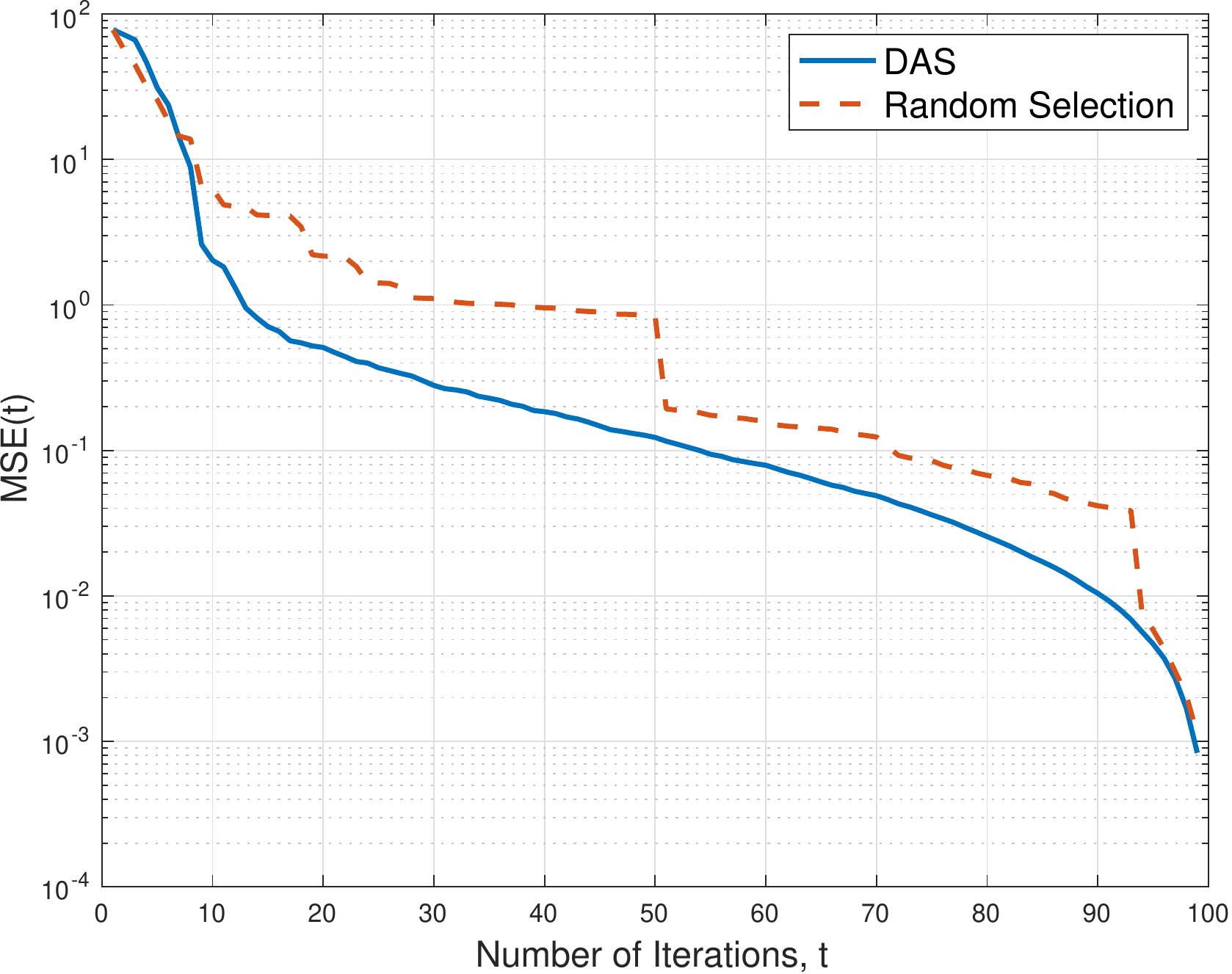} \\
(a) \\
\includegraphics[width=\figwidth]{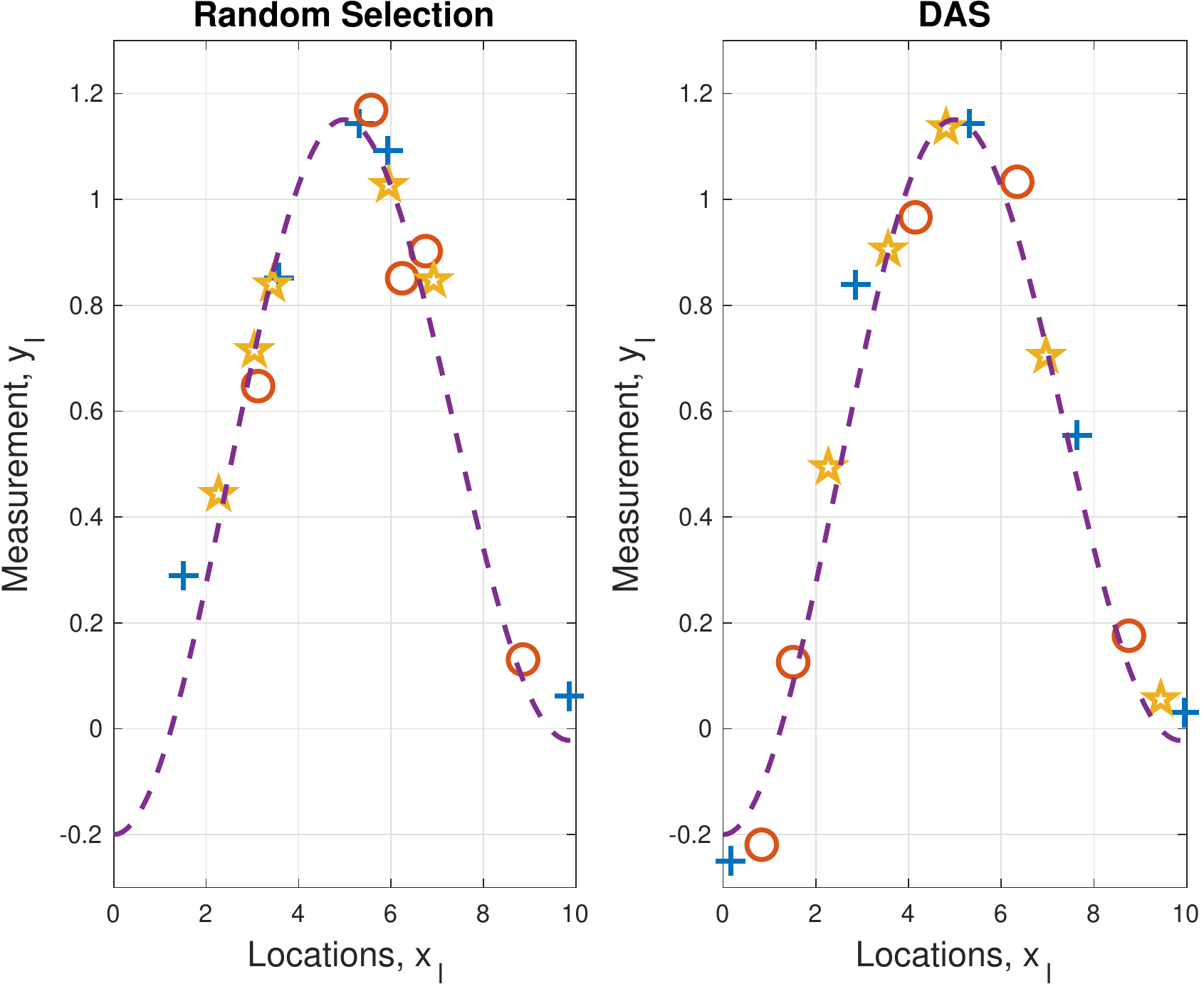} \\
(b) 
\end{center}
\caption{GPR results with $L = 100$ and $\sigma^2 = 0.01$: 
(a) the MSEs as functions of round or iteration;
(b) the uploaded measurements at the BS
(the first 5 
uploaded measurements
are represented by $+$ markers,
the next 5 by $\circ$ markers,
and the next 5 by $\star$ markers.}
        \label{Fig:plt1}
\end{figure}

As a toy example, we consider one-dimensional 
sensor network with $L = 100$ sensors with 
\be
m(x) = \left( \sin
\left( \frac{x}{3} \right)
\right)^2 - \frac{1}{5} \cos
\left( \frac{x}{2} \right), \ 0 \le x \le 10,
	\label{EQ:mx1}
\ee
and
$$
y_l = m(x_l) + n_l,
$$
where $n_l \sim \cN(0, \sigma^2)$ is the noise.
We assume that $x_l$ is 
random and uniformly distributed over $[0, 10]$.
As shown in Fig.~\ref{Fig:plt1}, $m(x)$
over $x \in [0, 10]$ is sufficiently smooth as the maximum
frequency of $m(x)$ in \eqref{EQ:mx1} 
is $f_{\rm max} = \frac{1}{3 \pi} \approx 0.1$.
In Fig.~\ref{Fig:plt1},
GPR results are shown with $L = 100$ 
sensors and $\sigma^2 = 0.01$. 
In particular, Fig.~\ref{Fig:plt1} (a)
shows that the MSE can rapidly decrease due to DAS
compared to random selection.
Thus, DAS allows GPR to provide
a good estimate of
the complete data set, $\cY$, with a small number of sensors uploading
measurements (e.g., 20 sensors)
compared to random selection,
while DAS and random selection provide the same MSE
after all the sensors upload their measurements.
The positions of sensors that upload
their measurements are shown
in Fig.~\ref{Fig:plt1} (b)
when random selection 
(on the left-hand side) and DAS 
(on the right-hand side) are used.
Note that in Fig.~\ref{Fig:plt1} (b),
the dashed line represents 
$m(x_l)$, while the markers stand for actual
noisy measurements at sensors.
Clearly, we can see that the positions of 
uploaded sensors are more uniformly distributed by
DAS than random selection,
which leads to a better GPR performance
using DAS (in terms of MSE as shown in
Fig.~\ref{Fig:plt1} (a)).

\begin{figure}[thb]
\begin{center}
\includegraphics[width=\figwidth]{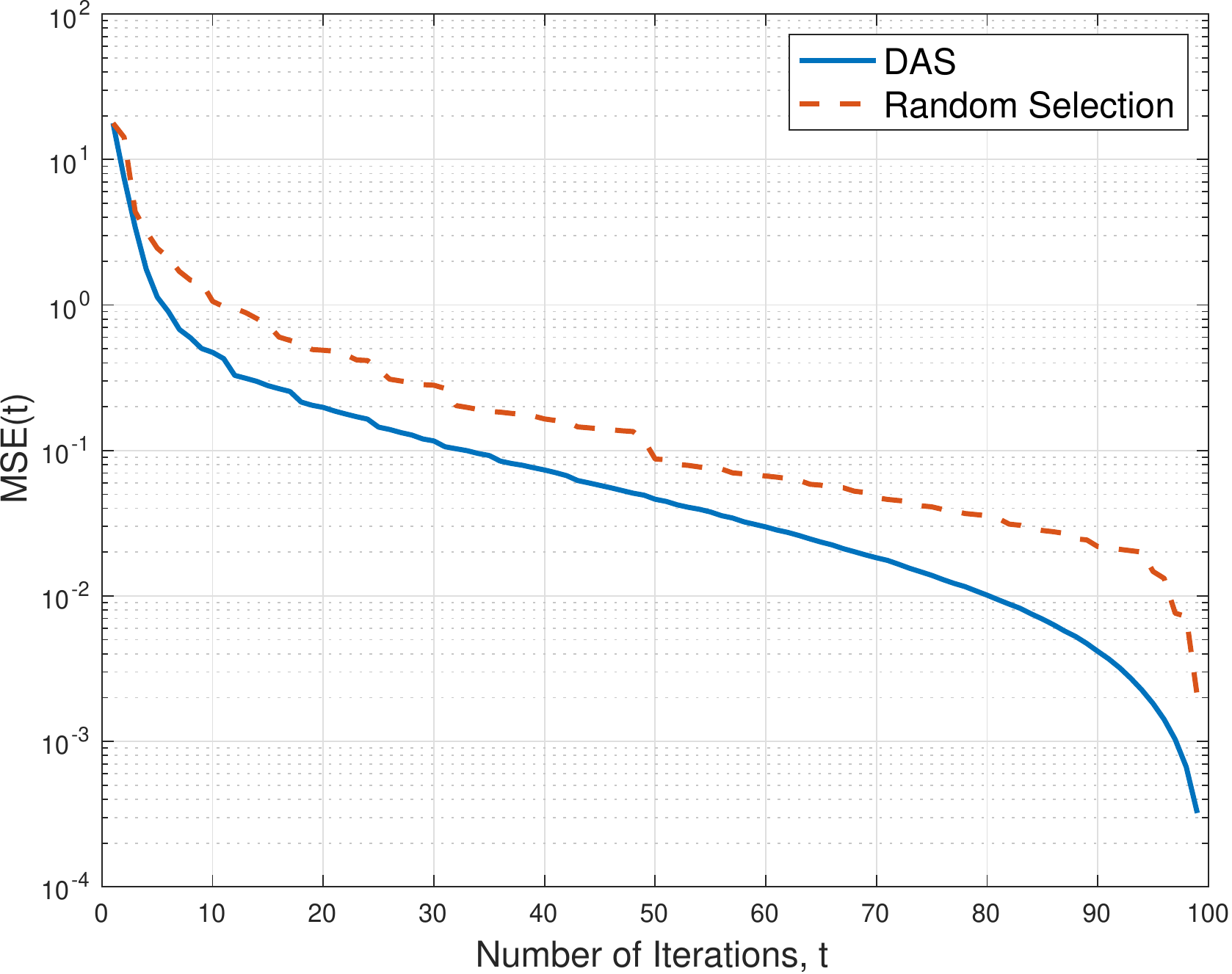} \\
(a) \\
\includegraphics[width=\figwidth]{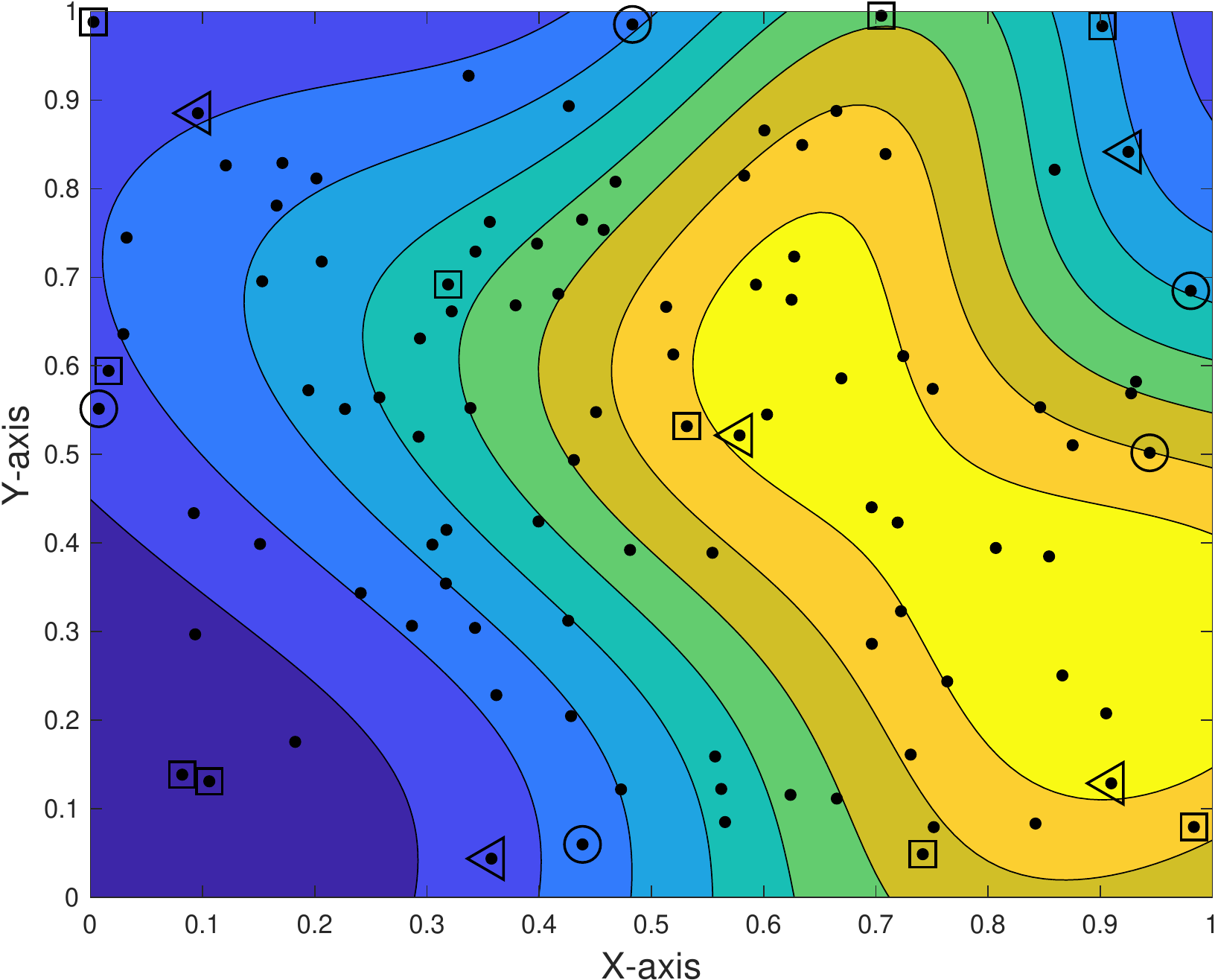} \\
(b) \\
\includegraphics[width=\figwidth]{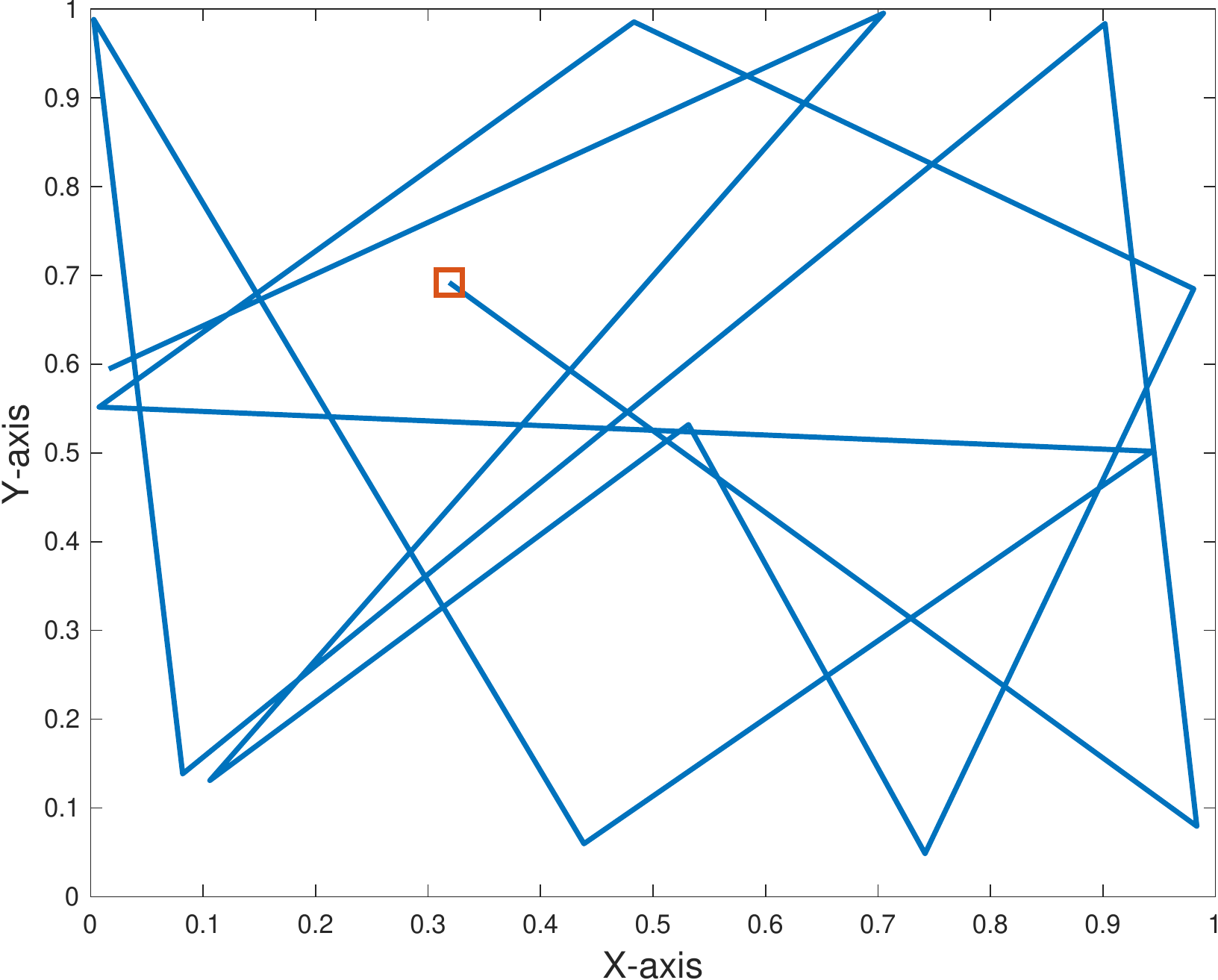} \\
(c) 
\end{center}
\caption{GPR results with $L = 100$ and $\sigma^2 = 0.1$:
(a) the MSEs as functions of round or iteration;
(b) a 2-dimensional view of $m(\bx)$ with the locations
of sensors (represented by dots)
and those of the sensors that uploaded measurements at the BS
(the first 5
uploaded measurements
are represented by $\square$ markers,
the next 5 by $\circ$ markers,
the next 5 by $\star$ markers, and
the next 5 by $\triangleleft$ markers);
(c) the trace of the first 15 uploaded sensors
in a run
(the $\square$ marker
represents the first sensor).}
        \label{Fig:plt2}
\end{figure}

For another toy example, consider a two-dimensional sensor
network 
with the following $m(\bx)$:  
\be
m(\bx) = 
\frac{1}{3} \sum_{i=1}^3 \exp
\left(- (\bx- \bp_i)^\rT \bA_i (\bx - \bp_i) \right),
\ \bx \in [0,1]^2,
	\label{EQ:mm2}
\ee
where $\bp_1 =[0.1 \ 0.9]^\rT$,
$\bp_2 =[0.5 \ 0.6]^\rT$,
$\bp_3 =[0.9 \ 0.7]^\rT$,
and
$$
\bA_1 = \left[ \begin{array}{cc}
4 & -6 \cr -1& 6 \cr
\end{array} \right],
\bA_2 = \left[ \begin{array}{cc}
8 & 1 \cr 5& 4 \cr
\end{array} \right],
\bA_3 = \left[ \begin{array}{cc}
8 & -4.1 \cr -4.1& 20 \cr
\end{array} \right].
$$
Fig.~\ref{Fig:plt2}
shows the performance of GPR with 
the 2-dimensional measurement model
in \eqref{EQ:mm2} when $L = 100$ and $\sigma^2 = 0.1$.
To see the performance, $L = 100$ locations are randomly generated
within an area of $[0,1]^2$ for sensors
(as shown by dots in Fig.~\ref{Fig:plt2} (b)).
The MSEs as functions of iteration or round are shown in
Fig.~\ref{Fig:plt2} (a), where
GPR with DAS can decrease the MSE rapidly
compared to GPR with random selection.
In Fig.~\ref{Fig:plt2} (b), 
the positions of sensors that upload
their measurements are shown.
Clearly, DAS can allow GPR to take measurements 
from more uniformly located sensors
and improve the performance of GPR.
We also show the trace of the positions of uploaded sensors
in a run in Fig.~\ref{Fig:plt2} (c), 
where the first sensor is shown by $\square$ marker.
It is shown that any two consecutive sensors 
are not close to each other. This is due to
the selection criterion in \eqref{EQ:lt1}.
That is, since the next sensor should have the largest
covariance (or uncertainty)
among those associated with $\cX_1(t)$,
any sensor that is close to the current one
cannot be chosen as it may have a small 
(conditional) covariance.

\begin{figure}[thb]
\begin{center}
\includegraphics[width=\figwidth]{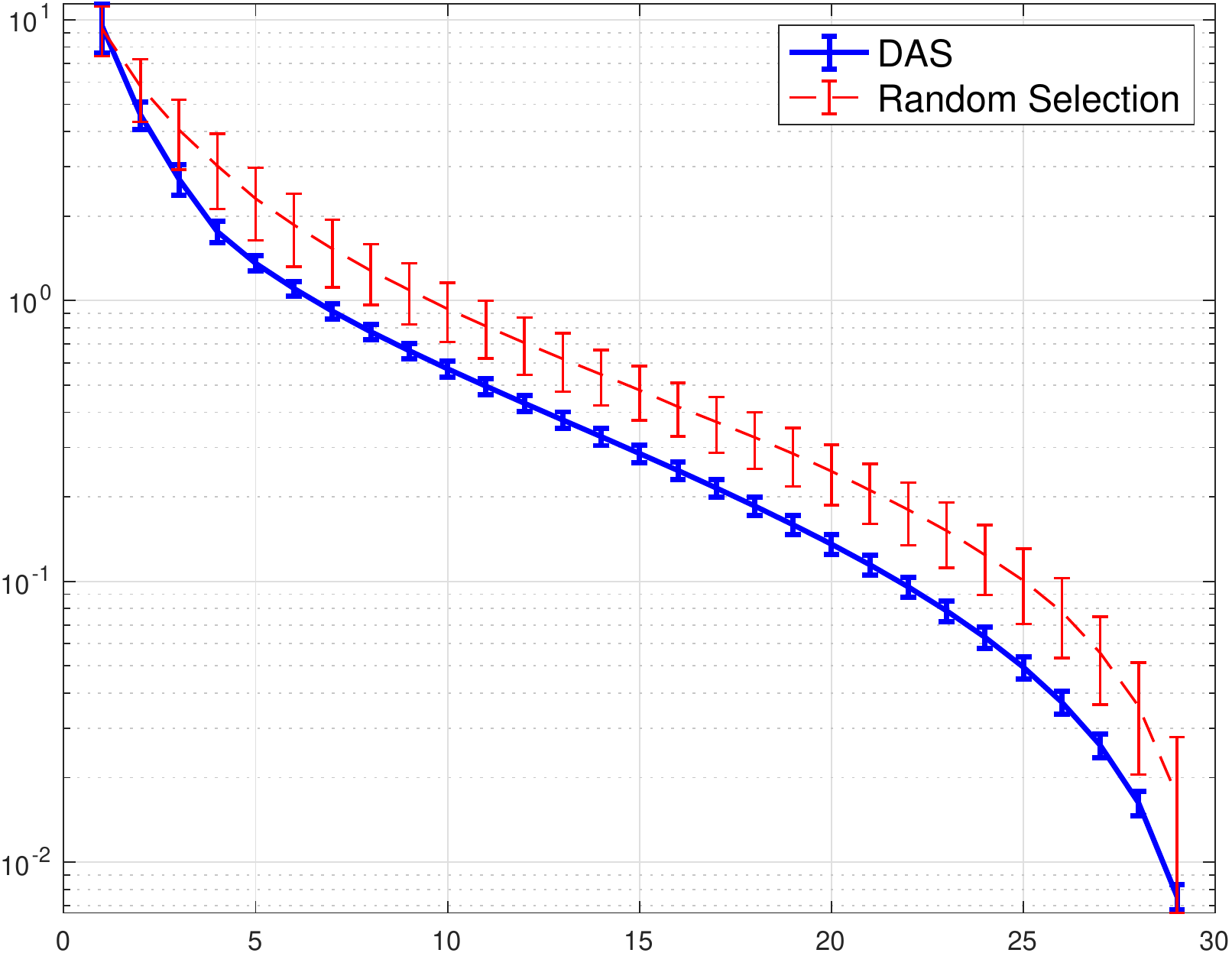} \\
\end{center}
\caption{The mean and standard deviation
of the MSEs with $L = 30$ and $\sigma^2 = 0.1$.
The vertical bars represent the standard deviations.}
        \label{Fig:plt_R1}
\end{figure}

The MSE curves in Figs.~\ref{Fig:plt1} -- \ref{Fig:plt2}
are the outcomes for single realization 
of $L$ random locations. 
To see the average performance,
we consider 1000 different realizations of $L$ locations,
where $L = 30$, and obtain the mean and standard deviation of MSEs.
The mean MSE curves are shown with standard deviations
in Fig.~\ref{Fig:plt_R1} with $\sigma^2 = 0.1$.
It is clearly shown that DAS can provide not only a lower MSE, but 
also a smaller standard deviation than random selection.

\begin{figure}[thb]
\begin{center}
\includegraphics[width=9cm]{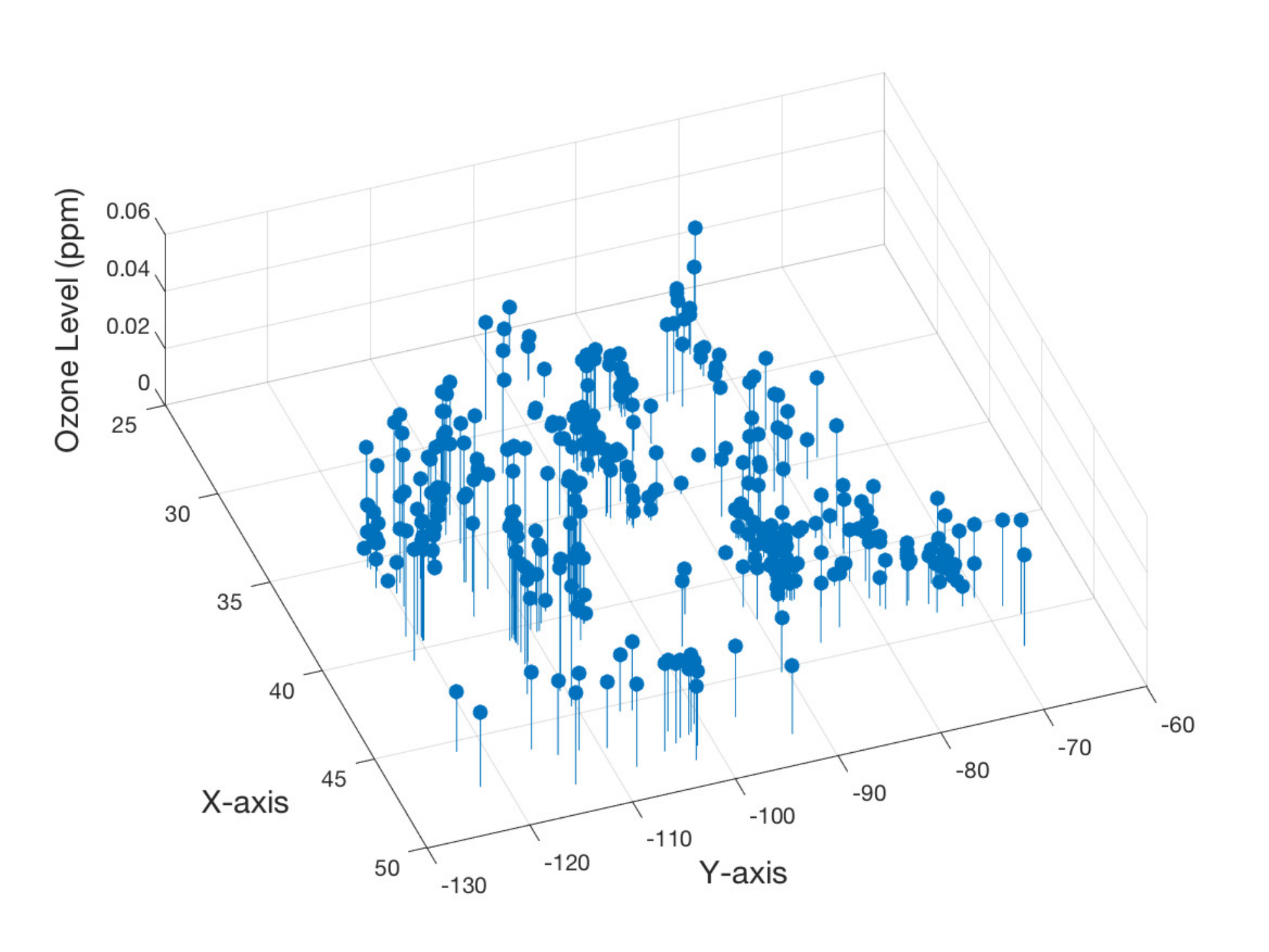} \\
(a) \\
\includegraphics[width=\figwidth]{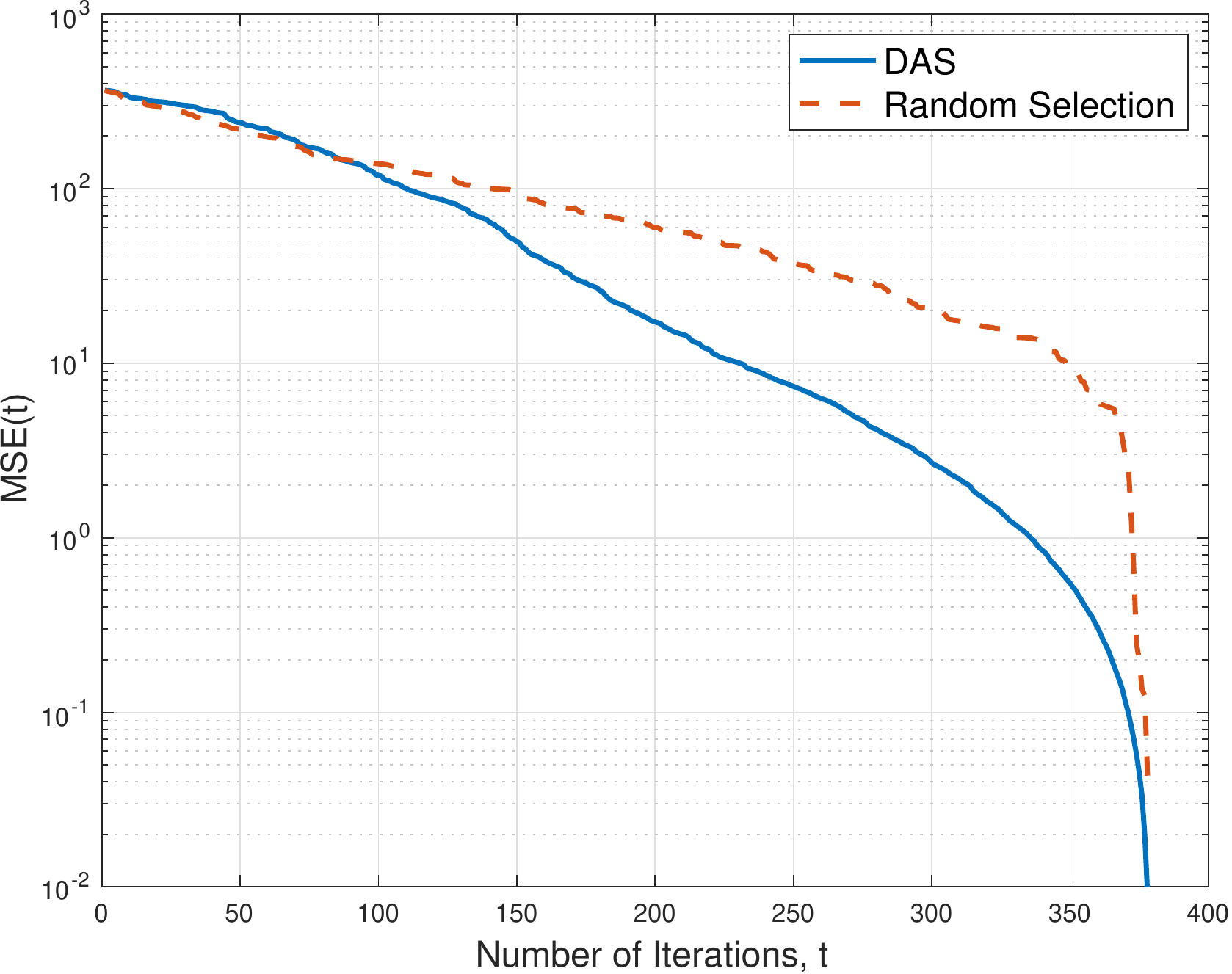} \\
(b)
\end{center}
\caption{GPR results with the ozone levels at
$L = 379$ locations:
(a) ozone levels in ppm at 379 locations;
(b) the MSEs as functions of round or iteration.}
        \label{Fig:real}
\end{figure}

In Figs.~\ref{Fig:plt1} -- \ref{Fig:plt_R1},
numerical results are provided with synthetic data sets.
GPR with DAS can also be applied to real data sets. 
A data set of hourly measured ozone levels at 379
stations that is available from
United State's Environmental Protection Agency (EPA)\footnote{Data
sets are available at 
``https://aqs.epa.gov."} is used.
In Fig.~\ref{Fig:real} (a),
the ozone levels (in parts per million (ppm)) at 379 locations
are shown from the data set.
The MSEs are shown in
Fig.~\ref{Fig:real} (b), which demonstrates
that GPR with DAS can decrease
the MSE than GPR with random selection. 
In other words, DAS allows GPR to learn efficiently with 
fewer measurements than random selection.
In particular, to achieve a target MSE of $10^{-2}$,
DAS requires about 220 measurements,
while random selection needs 350 measurements according to
Fig.~\ref{Fig:real} (b).
As a result, if there is a cost to obtain
each measurement, DAS can help lower
the cost in building a prediction model.

\subsection{Results with Multichannel ALOHA}

In this subsection, we consider
the case that $Q$ sensors are selected
at each round with $B$ channels.
Conventional multichannel ALOHA
with an equal probability of uploading,
$p_{\rm up} = \frac{B}{Q}$, is also considered
to compare with the modified multichannel ALOHA
where the predicted values of measurements
are fed back to the $Q$ selected sensors.
For simulations,
the following 1-dimensional model
is considered:
\be
m(x) = \sqrt\frac{2}{T} 
\sum_{i=1}^T 
X_i \sin \left(
2 \pi f_i x + \theta_i \right),
\ 0 \le x \le 10, 
\ee
where $X_i \sim \cN(0,1)$,
$f_i \sim {\rm Unif} (0, 1/2)$,
and $\theta_i \sim {\rm Unif} (0, 2 \pi)$.
For all simulations, we assume that $T = 10$
and in each run, $\{(X_i, f_i, \theta_i)\}$ 
is randomly generated. Each simulation result
is an average of 1000 runs.

To see the performance, 
the following SSE is used:
\be
\sSSE (\cU) = \sum_{q \in \cU} e_q^2,
\ee
where $\cU$ represents the set of the sensors
(among $Q$ selected
sensors) that fail to upload their measurements due to collision.
For convenience, at each round, $Q$ sensors are selected randomly 
rather than by any specific applications.
Note that a lower-bound on the SSE can be obtained as follows:
\be
\underline{\sSSE} = \sigma^2 (Q - B e^{-1}),
	\label{EQ:bSSE}
\ee
In \eqref{EQ:bSSE}, since $B e^{-1}$ is the average
number of the sensors that successfully upload
their measurement at each round (see \eqref{EQ:Smax}),
$Q-Be^{-1}$ is the average number of 
the sensors that fail to upload. In \eqref{EQ:e_q}, 
if the sensor's prediction is perfect,
$\uE[e_q^2]$ becomes $\sigma^2$, which leads to 
the lower-bound in \eqref{EQ:bSSE}.
That is, when GPR is able to perform good prediction
with a sufficient number of measurements
through a number of rounds,
the lower-bound in \eqref{EQ:bSSE} can be achieved.

Note that in 
the modified multichannel ALOHA with predictions, 
since sensors with large prediction errors
have high probabilities of uploading,
the sensors that do not upload more likely have
low prediction errors. As a result, due to the 
(distributed) selective uploading,
the SSE can be lower than that in \eqref{EQ:bSSE}
if the modified multichannel ALOHA with predictions
is used.

\begin{figure}[thb]
\begin{center}
\includegraphics[width=\figwidth]{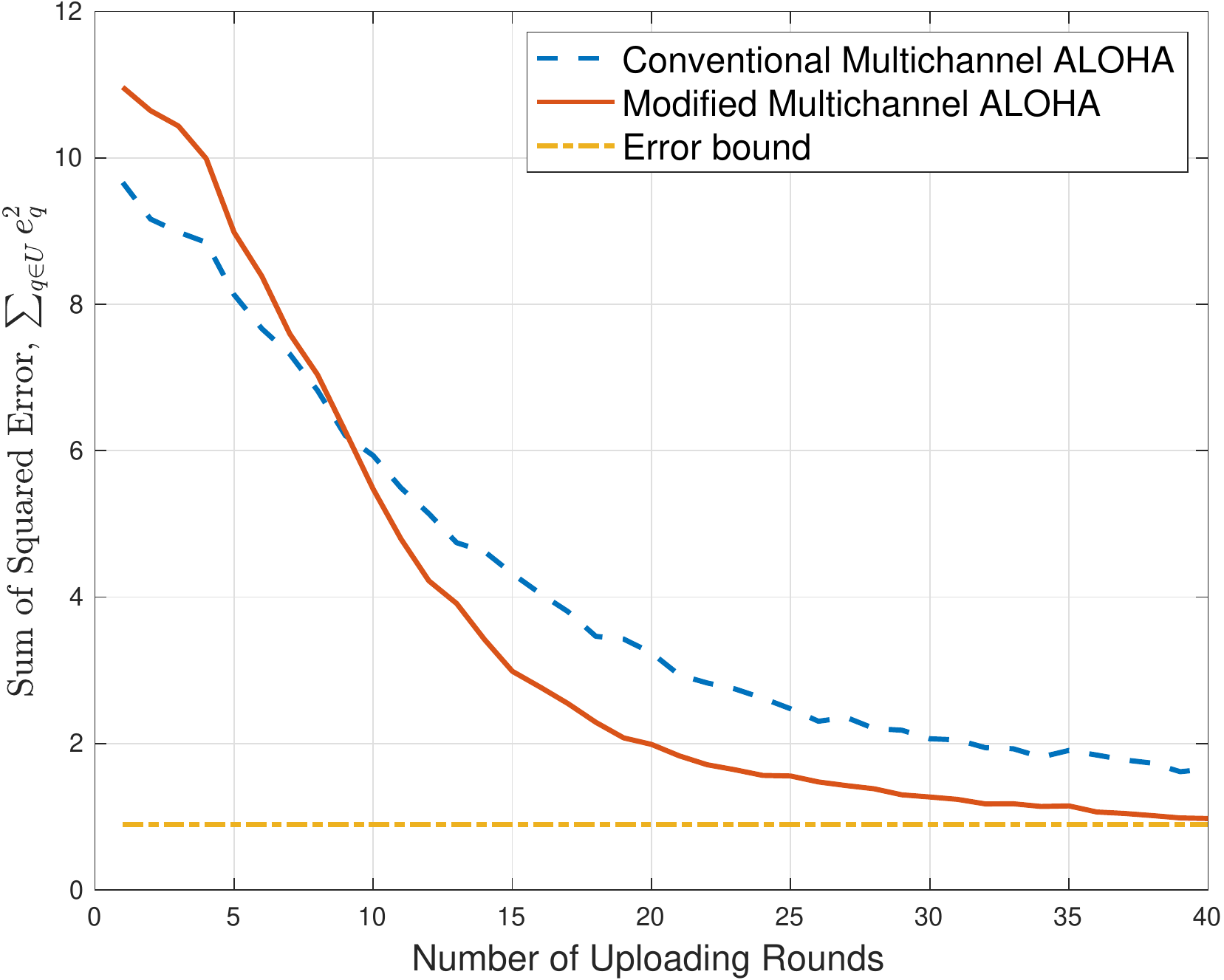} \\
\end{center}
\caption{The SSEs of conventional multichannel
ALOHA (with an equal probability of uploading,
$\frac{B}{Q}$) and modified multichannel ALOHA with $p_q$ in \eqref{EQ:p_ql} 
at each round with $L = 200$, $Q = 10$, $B = 3$, and $\sigma^2 = 0.1$.} 
\label{Fig:splt1}
\end{figure}

Fig.~\ref{Fig:splt1} shows 
the SSE at each round with $L = 200$, $Q = 10$,
$B = 3$, and $\sigma^2 = 0.1$.
For the case of conventional multichannel ALOHA,
the equal probability of uploading,
$p_{\rm up} = \frac{B}{Q}$, is used.
On the other hand, in the modified multichannel ALOHA,
$p_q$ is adaptively decided with
predicted values when the step size is given by $\mu = 0.5$.
It is shown that 
GPR with conventional multichannel ALOHA
can perform better than 
that with modified multichannel ALOHA
up to 8 or 9 rounds, although
modified multichannel ALOHA
can provide a better performance 
than conventional multichannel ALOHA for GPR
after 10 rounds.
This is mainly due to the time to adjust the
Lagrange multiplier
through the adaptive algorithm in \eqref{EQ:ll}.
It is also noteworthy that
after a sufficient number of rounds,
the SSE approaches the lower-bound in \eqref{EQ:bSSE}
as GPR can predict the measurements of the
sensors that do not upload yet.

\begin{figure}[thb]
\begin{center}
\includegraphics[width=\figwidth]{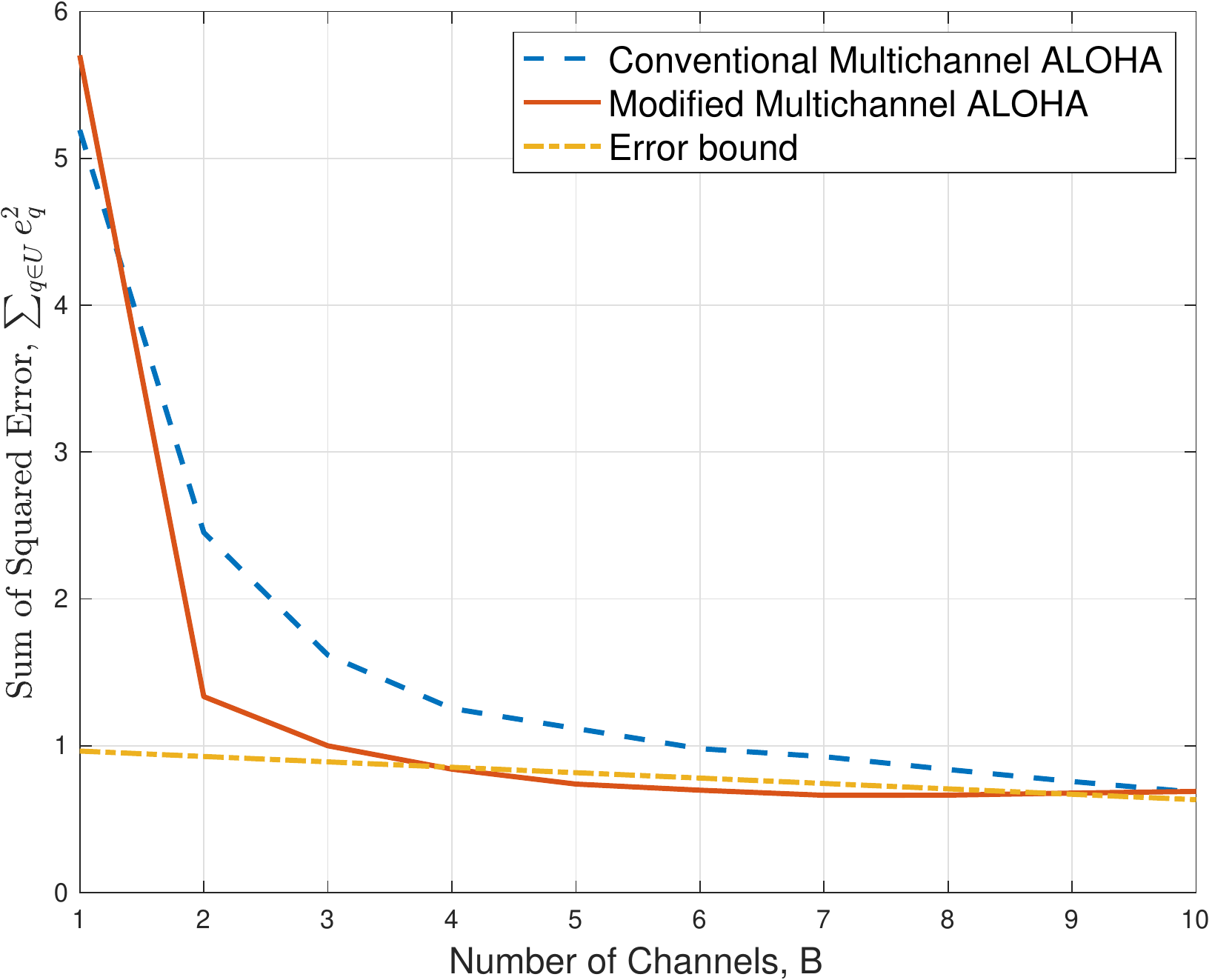} \\
\end{center}
\caption{The SSEs of conventional multichannel
ALOHA (with an equal probability of uploading,
$\frac{B}{Q}$) and modified multichannel ALOHA with $p_q$ in \eqref{EQ:p_ql} 
after 40 rounds as functions of $B$
with $L = 200$, $Q = 10$, and $\sigma^2 = 0.1$.} 
        \label{Fig:splt2}
\end{figure}

Fig.~\ref{Fig:splt2}
shows the SSEs of conventional multichannel
ALOHA (with an equal probability of uploading,
$\frac{B}{Q}$) and modified multichannel ALOHA with $p_q$ in \eqref{EQ:p_ql} 
after 40 rounds as functions of $B$
with $L = 200$, $Q = 10$, and $\sigma^2 = 0.1$.
For a fixed $Q$, at each round, there are more
sensors that succeed to upload their
measurements for a larger $B$ (i.e., more channels),
which leads to the decrease of SSE.
With modified multichannel ALOHA,
it is shown that $B = 2$ or 3 channels are sufficient
to approach the lower-bound on SSE.
We also note that the SSE of
modified multichannel ALOHA 
can be lower than the lower-bound on SSE,
which was explained earlier due to the selective
uploading.

\begin{figure}[thb]
\begin{center}
\includegraphics[width=\figwidth]{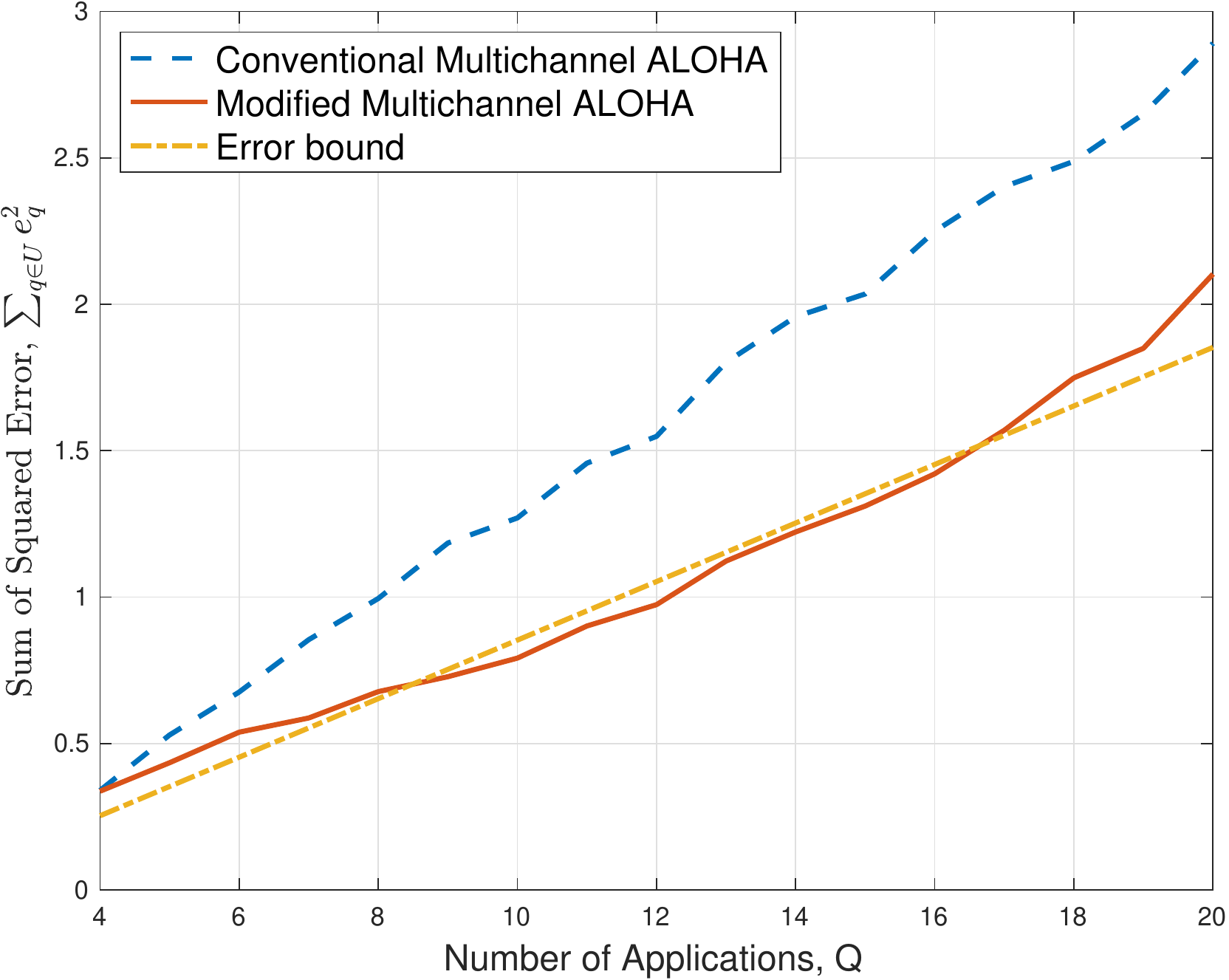} \\
\end{center}
\caption{The SSEs of conventional multichannel
ALOHA (with an equal probability of uploading,
$\frac{B}{Q}$) and modified multichannel ALOHA with $p_q$ in \eqref{EQ:p_ql} 
after 40 rounds as functions of $Q$
with $L = 200$, $B = 4$, and $\sigma^2 = 0.1$.} 
        \label{Fig:splt3}
\end{figure}

In Fig.~\ref{Fig:splt3},
the SSEs of conventional multichannel
ALOHA (with an equal probability of uploading,
$\frac{B}{Q}$) and modified multichannel ALOHA with $p_q$ in \eqref{EQ:p_ql} 
after 40 rounds are
shown as functions of $Q$
with $L = 200$, $B = 4$, and $\sigma^2 = 0.1$.
Since $B$ is fixed,
it is shown that the SSE increases with $Q$.
It is also observed that
GPR with modified multichannel ALOHA 
outperforms 
GPR with conventional multichannel ALOHA for 
a wide range of $Q$.

\section{Concluding Remarks}	\label{S:Con}

In this paper, 
GPR has been considered for IoT systems to 
learn data sets collected from sensors without any specific model
for data sets.
To efficiently collect measurements from sensors,
DAS was applied to GPR. In particular, we considered
the interpolation of sensors' measurements
from a small number of measurements 
uploaded by a fraction of sensors.
It was shown that GPR with DAS
can have a good estimate of a complete data set 
using measurements uploaded by 
a fraction of sensors thanks to active sensor selection
compared to GPR with random selection.
As a result, when the number of sensors
to upload their measurements is limited
(due to various reasons including uploading time constraints),
DAS can help GPR to efficiently learn data sets.
DAS was also generalized with multichannel ALOHA
where distributed selective uploading is employed,
since each selected sensor can decide whether or not it uploads
by comparing its measurement with the predicted one
that is fed back by the BS.

The approach in this paper could be seen
as an example to demonstrate how ML algorithms can learn data
sets with efficient sensing and communication schemes
(e.g., DAS). Thus, the approach can be extended
for classification (when sensors' measurements are to be labeled)
and inference, which would be further research topics to be studied
in the future.

\bibliographystyle{ieeetr}
\bibliography{sensor}

\begin{thebibliography}{10}

\bibitem{Gubbi13}
J.~Gubbi, R.~Buyya, S.~Marusic, and M.~Palaniswami, ``Internet of things
  ({IoT}): A vision, architectural elements, and future directions,'' {\em
  Future Gener. Comput. Syst.}, vol.~29, pp.~1645--1660, Sept. 2013.

\bibitem{Kim16}
J.~{Kim}, J.~{Yun}, S.~{Choi}, D.~N. {Seed}, G.~{Lu}, M.~{Bauer},
  A.~{Al-Hezmi}, K.~{Campowsky}, and J.~{Song}, ``Standard-based {IoT}
  platforms interworking: implementation, experiences, and lessons learned,''
  {\em IEEE Communications Magazine}, vol.~54, pp.~48--54, July 2016.

\bibitem{Fuqaha15}
A.~Al-Fuqaha, M.~Guizani, M.~Mohammadi, M.~Aledhari, and M.~Ayyash, ``Internet
  of {T}hings: A survey on enabling technologies, protocols, and
  applications,'' {\em IEEE Communications Surveys Tutorials}, vol.~17,
  pp.~2347--2376, Fourthquarter 2015.

\bibitem{Yaq17}
I.~{Yaqoob}, E.~{Ahmed}, I.~A.~T. {Hashem}, A.~I.~A. {Ahmed}, A.~{Gani},
  M.~{Imran}, and M.~{Guizani}, ``{I}nternet of {T}hings architecture: Recent
  advances, taxonomy, requirements, and open challenges,'' {\em IEEE Wireless
  Communications}, vol.~24, no.~3, pp.~10--16, 2017.

\bibitem{Ding_20Access}
J.~{Ding}, M.~{Nemati}, C.~{Ranaweera}, and J.~{Choi}, ``{IoT} connectivity
  technologies and applications: A survey,'' {\em IEEE Access}, vol.~8,
  pp.~67646--67673, 2020.

\bibitem{Mang16}
N.~{Mangalvedhe}, R.~{Ratasuk}, and A.~{Ghosh}, ``{NB-IoT} deployment study for
  low power wide area cellular {IoT},'' in {\em 2016 IEEE 27th Annual
  International Symposium on Personal, Indoor, and Mobile Radio Communications
  (PIMRC)}, pp.~1--6, Sep. 2016.

\bibitem{3GPP_NBIoT}
3GPP TS 36.321 V13.2.0, {\em Evolved Universal Terrestrial Radio Access
  ({E-UTRA}); Medium Access Control ({MAC}) protocol specification}, June 2016.

\bibitem{Soua11}
R.~{Soua} and P.~{Minet}, ``A survey on energy efficient techniques in wireless
  sensor networks,'' in {\em 2011 4th Joint IFIP Wireless and Mobile Networking
  Conference (WMNC 2011)}, pp.~1--9, 2011.

\bibitem{Abella19}
C.~S. {Abella}, S.~{Bonina}, A.~{Cucuccio}, S.~{D’Angelo}, G.~{Giustolisi},
  A.~D. {Grasso}, A.~{Imbruglia}, G.~S. {Mauro}, G.~A.~M. {Nastasi},
  G.~{Palumbo}, S.~{Pennisi}, G.~{Sorbello}, and A.~{Scuderi}, ``Autonomous
  energy-efficient wireless sensor network platform for home/office
  automation,'' {\em IEEE Sensors Journal}, vol.~19, no.~9, pp.~3501--3512,
  2019.

\bibitem{Bhuiyan13}
M.~Z.~A. {Bhuiyan}, G.~{Wang}, J.~{Cao}, and J.~{Wu}, ``Energy and
  bandwidth-efficient wireless sensor networks for monitoring high-frequency
  events,'' in {\em 2013 IEEE International Conference on Sensing,
  Communications and Networking (SECON)}, pp.~194--202, 2013.

\bibitem{Bockelmann18}
C.~{Bockelmann}, N.~K. {Pratas}, G.~{Wunder}, S.~{Saur}, M.~{Navarro},
  D.~{Gregoratti}, G.~{Vivier}, E.~{De Carvalho}, Y.~{Ji}, C.~{Stefanović},
  P.~{Popovski}, Q.~{Wang}, M.~{Schellmann}, E.~{Kosmatos}, P.~{Demestichas},
  M.~{Raceala-Motoc}, P.~{Jung}, S.~{Stanczak}, and A.~{Dekorsy}, ``Towards
  massive connectivity support for scalable {mMTC} communications in {5G}
  networks,'' {\em IEEE Access}, vol.~6, pp.~28969--28992, 2018.

\bibitem{Choi19}
J.~{Choi}, ``A cross-layer approach to data-aided sensing using compressive
  random access,'' {\em IEEE Internet of Things J.}, vol.~6, pp.~7093--7102,
  Aug 2019.

\bibitem{Choi20_IoT}
J.~{Choi}, ``Gaussian data-aided sensing with multichannel random access and
  model selection,'' {\em IEEE Internet of Things J.}, vol.~7, no.~3,
  pp.~2412--2420, 2020.

\bibitem{Williams96}
C.~K.~I. Williams and C.~E. Rasmussen, ``Gaussian processes for regression,''
  in {\em Advances in Neural Information Processing Systems 8}, pp.~514--520,
  MIT press, 1996.

\bibitem{Rasm06}
C.~E. Rasmussen and C.~K.~I. Williams, {\em Gaussian Processes for Machine
  Learning}.
\newblock Cambridge, MA, USA: MIT Press, 2006.

\bibitem{Bishop06}
C.~M. Bishop, {\em Pattern Recognition and Machine Learning}.
\newblock New York: Springer, 2006.

\bibitem{Wang20}
J.~{Wang}, C.~{Jiang}, H.~{Zhang}, Y.~{Ren}, K.~C. {Chen}, and L.~{Hanzo},
  ``Thirty years of machine learning: The road to {P}areto-optimal wireless
  networks,'' {\em IEEE Communications Surveys Tutorials}, vol.~22, no.~3,
  pp.~1472--1514, 2020.

\bibitem{Cohn96}
D.~A. Cohn, Z.~Ghahramani, and M.~I. Jordan, ``Active learning with statistical
  models,'' {\em J. Artif. Int. Res.}, vol.~4, p.~129–145, Mar. 1996.

\bibitem{Shen03}
D.~Shen and V.~O.~K. Li, ``Performance analysis for a stabilized multi-channel
  slotted {ALOHA} algorithm,'' in {\em Proc. IEEE PIMRC}, vol.~1, pp.~249--253
  Vol.1, Sept 2003.

\bibitem{Chang15}
C.~H. Chang and R.~Y. Chang, ``Design and analysis of multichannel slotted
  {ALOHA} for machine-to-machine communication,'' in {\em Proc. IEEE GLOBECOM},
  pp.~1--6, Dec 2015.

\bibitem{Jurdak10}
R.~{Jurdak}, A.~G. {Ruzzelli}, and G.~M.~P. {O'Hare}, ``Radio sleep mode
  optimization in wireless sensor networks,'' {\em IEEE Trans. Mobile
  Computing}, vol.~9, no.~7, pp.~955--968, 2010.

\bibitem{Galinina13}
O.~Galinina, A.~Turlikov, S.~Andreev, and Y.~Koucheryavy, ``Stabilizing
  multi-channel slotted {ALOHA} for machine-type communications,'' in {\em
  Proc. IEEE ISIT}, pp.~2119--2123, July 2013.

\bibitem{Choi16CL}
J.~Choi, ``On the adaptive determination of the number of preambles in {RACH}
  for {MTC},'' {\em IEEE Communications Letters}, vol.~20, pp.~1385--1388, July
  2016.

\bibitem{Boyd11}
S.~Boyd, N.~Parikh, E.~Chu, B.~Peleato, and J.~Eckstein, ``Distributed
  optimization and statistical learning via the alternating direction method of
  multipliers,'' {\em Found. Trends Mach. Learn.}, vol.~3, pp.~1--122, Jan.
  2011.

\bibitem{Choi20_WCL}
J.~{Choi} and S.~R. {Pokhrel}, ``Federated learning with multichannel
  {ALOHA},'' {\em IEEE Wireless Communications Letters}, vol.~9, no.~4,
  pp.~499--502, 2020.

\end{thebibliography}

\end{document}